\documentclass[%
 tightenlines,
 nobibnotes,
 amsmath,amssymb,
 aps,
 rmp,
 11pt,a4paper
]{revtex4-1}

\usepackage{helvet}         
\usepackage{courier}        
\usepackage{type1cm}        
\usepackage{natbib}
\usepackage{graphicx} 
\usepackage[bottom]{footmisc}

\begin{document}

\title{Phonon-assisted radiofrequency absorption 
by gold nanoparticles resulting in hyperthermia}
\author{Andrei Postnikov}
\affiliation{Universit\'e de Lorraine, LCP-A2MC, 1 Bd Arago, F-57078 Metz, France}
\email{andrei.postnikov@univ-lorraine.fr}
\author{Kamil Moldosanov}
\affiliation{Kyrgyz-Russian Slavic University, 44 Kiyevskaya St., Bishkek 720000, Kyrgyzstan}
\email{altair1964@yandex.ru}

\begin{abstract}
It is suggested that in gold nanoparticles (GNPs) of about 5~nm sizes
used in the radiofrequency (RF) hyperthermia, an absorption of the RF photon
by the Fermi electron occurs with involvement of the longitudinal acoustic
vibrational mode (LAVM), the dominating one in the distribution of vibrational
density of states (VDOS). This physical mechanism helps to explain two observed
phenomena: the size dependence of the heating rate (HR) in GNPs
and reduced heat production
in aggregated GNPs. The argumentation proceeds within the one-electron approximation,
taking into account the discretenesses of energies and momenta of both electrons
and LAVMs. The heating of GNPs is thought to consist of two consecutive processes:
first, the Fermi electron absorbs simultaneously the RF photon and the LAVM
available in the GNP; hereafter the excited electron gets relaxed within the GNP's
boundary, exciting a LAVM with the energy higher than that of the previously
absorbed LAVM. GNPs containing the Ta and/or Fe impurities are proposed for
the RF hyperthermia as promising heaters with enhanced HRs, and GNPs with
rare-earth impurity atoms are also brought into consideration. It is shown
why the maximum HR values should be expected in GNPs with about 
$5 - 7$~nm size.

\vspace*{10mm}
\includegraphics[width=8.0cm]{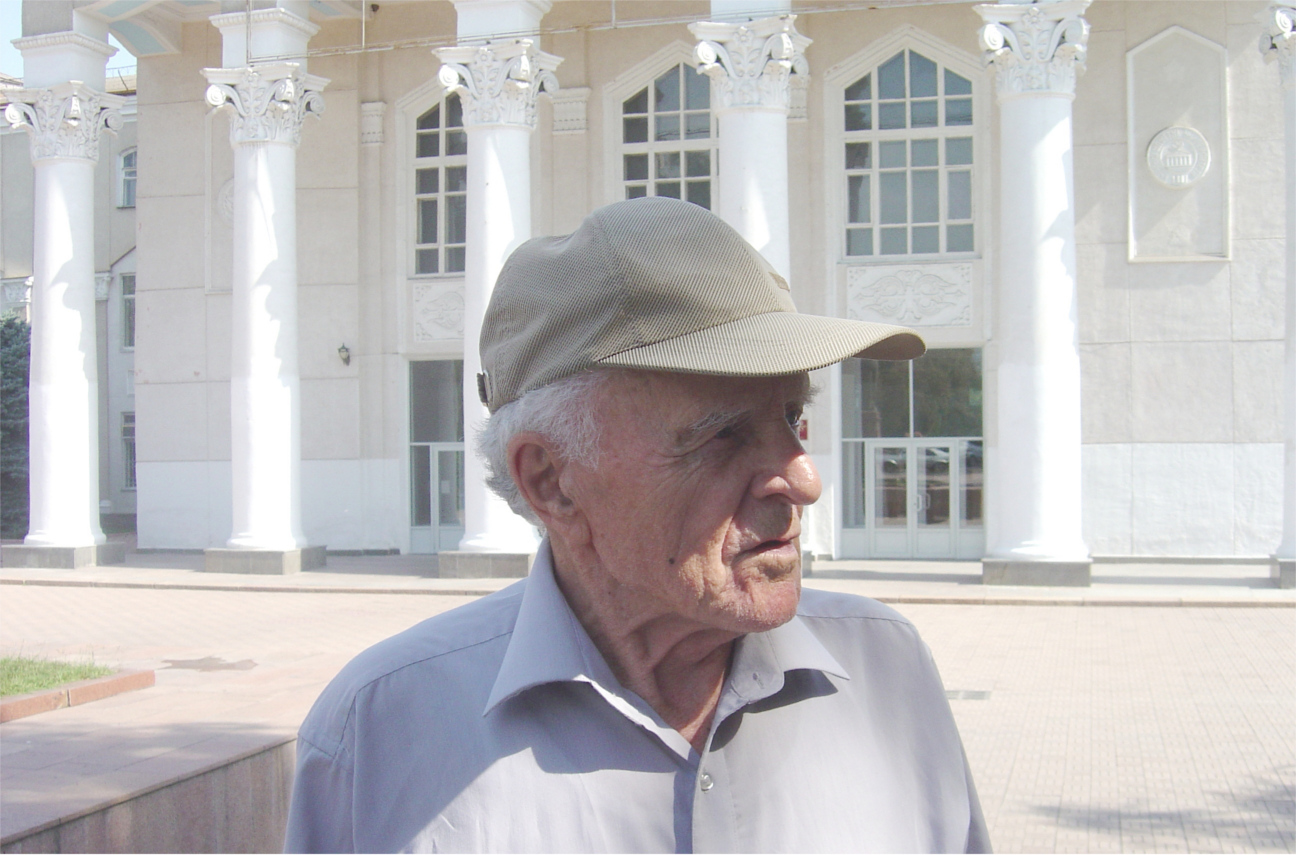}

\vspace*{1mm}\hspace*{2mm}
\parbox[c]{9.0cm}{\sf
The authors dedicate this work to the 90th anniversary of\\
professor Leonid V. Tuzov of the Kyrgyz State University}
\end{abstract}

\maketitle

\section{Introduction}
\label{sec:1}
Among methods allowing for hyperthermia in biological tissues, that based
on the radiofrequency (RF) absorption by gold nanoparticles (GNPs)
got in the last years much attention, due to its important advantage --
a deep penetration of the RF radiation into biological tissue 
($\sim$30~cm at frequencies $\sim$10~MHz). Details and useful references
can be found in the review by \citet{Corr2012chapter}. 
\citet{NanoRes2-400} in their experiments
carried out at 13.56~MHz revealed that
the heating rate (HR) depends on the particle size, namely that smaller particles
are heated faster, presumably because of their higher resistivity. However,
the exact physical basis of heat generation by nanoparticles remained so far
unclarified. This is a serious obstacle for grasping the HR / particle size relation
and hence for making ``intelligent choice'' of the GNP size, tailored
to the particular use. For example, one may wish to assure an enhanced
specific absorption rate (SAR) to heat the GNP to necessary temperatures
at minimum released power, avoiding at the same time an unnecessary heating
of normal tissues during hyperthermia.

\begin{figure}[b]
\noindent
\parbox[b]{9.8cm}{%
\includegraphics[width=9.0cm]{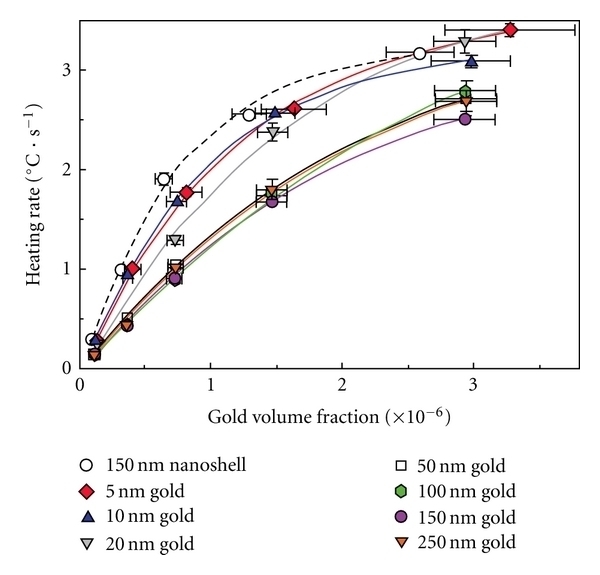}
}
\parbox[b]{6.0cm}{
\caption{Size-dependent heating of gold nanoparticles (diameters indicated)
in the RF field of the frequency 13.56~MHz. Reproduced with permission
from \citet{NanoRes2-400}.}
\label{fig_01}
}   
\end{figure}

According to standard definition, SAR = $C({\Delta}T/{\Delta}t)$, where $C$ is
the specific heat and ${\Delta}T/{\Delta}t$ is the HR (temperature / time).
The size dependence of the HR for the GNPs used by \citet{NanoRes2-400}
is presented in Fig.~\ref{fig_01}.
As is seen, the HR attains maximal values, among the nanoparticles studied,
in the smallest ones, of the 5~nm size. Along with \citet{NanoRes2-400}, also 
the observations of \citet{IntJHepato2011-676957,IEEE_TransBiomedEng58-2002}
imply that the smaller the size of GNPs (for a given volume fraction of gold),
the higher their HR: among experimentally studied GNPs with diverse sizes,
those of $\sim$5~nm diameter exhibit the highest one.
\citet{JPhysChemC116-24380} strengthened this conclusion,
having stated that the GNPs only of size 
10~nm or less could be heated by the electromagnetic RF field (again, the highest 
experimentally observed HR corresponded to the GNPs of 5~nm diameter).

In this paper, a physical model of the size effect in heat generation and reduction
of heat generation in aggregated GNPs is proposed, in which the longitudinal acoustic
vibrational modes (LAVMs) play an important role. None of the theoretical models
reviewed critically by \citet{Corr2012chapter} took these latter into account,
whereas
we will show that the inclusion of acoustic modes enables to explain a number
of experimental results. Besides, our findings make it possible to obtain
an optimal size of GNPs providing maximum SAR. 

The manuscript is organised as follows. Sec.~\ref{sec:2} shows that in large GNPs
(of the diameter exceeding 162~nm) the heating by RF radiation is possible owing to
uncertainty in the Fermi electron's momentum; however, the use of this mechanism
in hyperthermia cannot be but very limited. For practical treatment,
much smaller GNPs need to be -- and indeed are -- used, revealing
a pronounced size dependence of the heating efficiency. Possible physical
foundations of the related heating mechanism, along with explanation of
the observed size effect, are discussed in Sec.~\ref{sec:3}.
In Sec.~\ref{sec:4}, on the basis of conclusions drawn from Sec.~\ref{sec:3},
the experimentally observed reduced heat production in aggregated GNPs is explained. 
Sec.~\ref{sec:5} addresses possible ways of further
increase of the HP in GNPs. The general discussion of results is offered
in Sec.~\ref{sec:6}, and the conclusion 
drawn in Sec.~\ref{sec:7}.

\section{Setting the stage; conditions for the RF absorption}
\label{sec:2}
First we discuss an issue of whether the phonons are at all needed
for the RF absorption, and what are the conditions for the latter to happen
in a ``conventional'' way. The key feature in absorption of electromagnetic waves
by nanoparticles is quantization of energy levels, which become denser
as the particle size increases. Satisfying the conditions of conservation
of both energy and momentum on absorption is not, \emph{a priori}, possible
without additional considerations, because the curvature/slope of dispersion
relations are very different for electrons and photons. We consider first
the case when no other players enter the picture, and assume the matching conditions
to be satisfied within the uncertainty relation. To be specific, we take the GNP
to be spherical, of diameter $D$; the RF of 13.56~MHz, like in the above experiment,
corresponds to photon energy $h\nu = 5.6{\cdot}10^{-8}$~eV and momentum
$p_{\rm ph}\approx 3{\cdot}10^{-30}$~g$\cdot$cm$\cdot$s$^{-1}$.

The skin depth in gold at 13.56~MHz is about 20~$\mu$m, i.e., by orders of magnitude
exceeding the GNP size. One can therefore assume that the electric field
penetrates the volume and is of the same strength throughout the particle.

In metal particles as small as $\sim$5~nm, and even for much larger ones,
the quantization (discreteness) of the electron energy spectrum
is a crucial factor shaping their properties. Speaking specifically about gold,
one can advance far enough on the basis of the free-electron model,
assuming one electron per atom and estimating the corresponding Fermi energy. 

\begin{figure}[b]
\parbox[b]{8.4cm}{%
\includegraphics[width=7.6cm]{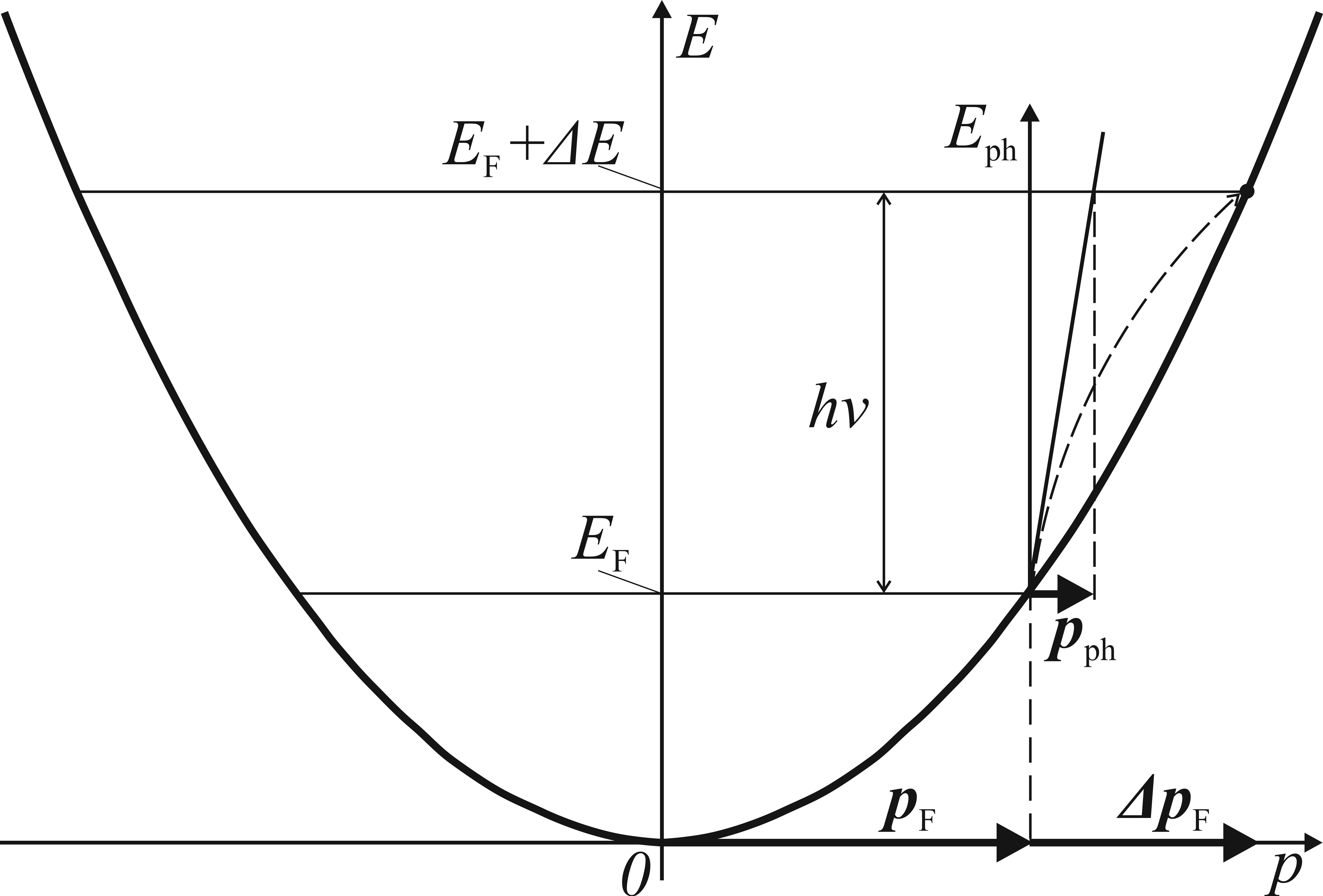}}
\parbox[b]{7.0cm}{%
\caption{\small
Scheme of absorption of RF photon by the Fermi electron without involvement of LAVM.
Dashed curve indicates a transition of the Fermi electron excited by the RF photon
with energy $h\nu$.}
\label{fig_02}
}    
\end{figure}

Assume that the absorption happens due to transitions between discrete levels
induced by quantum confinement; the free-electron dispersion relation
$E = p^2/2m$ implies  that the energy step $\Delta E = E_{\rm ph}$
must be accompanied by the correction of momentum, 
$\Delta p_{\rm F}\approx E_{\rm ph} m/p_{\rm F} = h\nu/v_{\rm F}$
(see Fig.~\ref{fig_02}).
For the relevant values of $h\nu = 5.6{\cdot}10^{-8}$~eV and the Fermi velocity
of gold as in the free-electron model
($v_{\rm F}\approx1.4{\cdot}10^6$~m$\cdot$s$^{-1}$),
the momentum mismatch amounts to 
$\Delta p_{\rm F}\approx 6.4{\cdot}10^{-28}$~g$\cdot$cm$\cdot$s$^{-1}$,
two orders of magnitude larger than what the photon momentum (see above)
could bring about. However, the small size of GNP may result in appreciable 
Heisenberg's uncertainty of electron momenta in them and thus ``smear out''
the discussed mismatch condition. Let us make some estimates to this end.

Following the pioneering works by \cite{JPSJ17-975,JPhysColloq38-C2-69},
we remind a relationship between the diameter $D$ and the level spacing
${\Delta}E$. Assume that the electron energy level $E'$ is the
first vacant one above the Fermi level $E_{\rm F}$. 
The level spacing
${\Delta}E_{\rm el} = E'-E_{\rm F}$ and the number of atoms $N_a$ in the GNP
are related as follows \citep[the derivation of the Kubo's formula
is given in Appendix \ref{sec:App1}]{JPSJ17-975}:
$$
\Delta E_{\rm el}\approx (4/3){\cdot}E_{\rm F}/N_a\,,
\quad\mbox{or}\quad
N_a \approx (4/3){\cdot}E_{\rm F}/\Delta E_{\rm el}\,;
$$

\vspace*{-6mm}

\begin{equation} 
D \approx \left(
\frac{8}{\pi}\,\frac{m_a}{\rho}\,\frac{E_{\rm F}}{\Delta E_{\rm el}}
\right)^{\!\!1/3}\,,
\label{eq:D-a}
\end{equation} 

\begin{figure}[t]
\includegraphics[width=14.0cm]{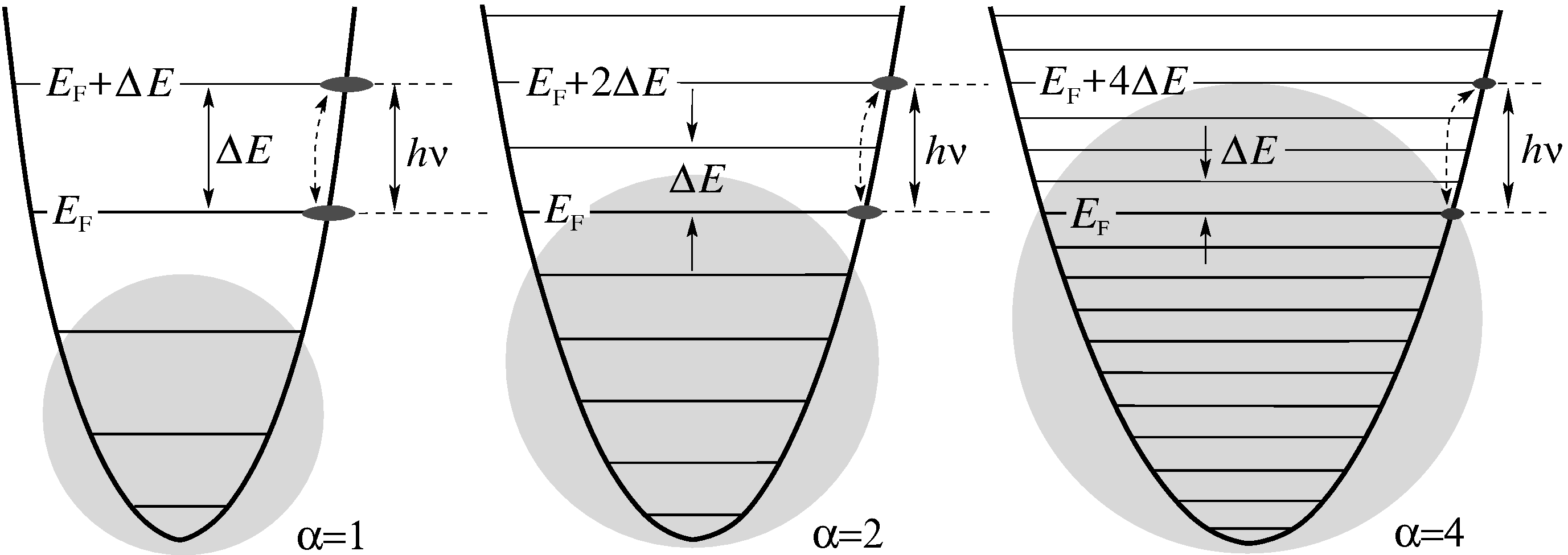}
\caption{\small
Schemes of ``direct'' absorption of RF photons $h\nu$ by electrons
at the Fermi level $E_{\rm F}$, for different energy discretisation steps
${\Delta}E$, depending on the particle size, and for different step 
multiplicities $\alpha$ (see text and Table~\ref{tab:1} for details).
The uncertainty of the electron's momentum is hinted by elliptical dots
of different width.} 
\label{fig_03}       
\end{figure}

\noindent
where $m_a$ is the atomic mass of gold, 
$m_a = 197$ in atomic units, or $\approx\,3.27{\cdot}10^{-22}$~g;
$\rho=19.3$~g$\cdot$cm$^{-3}$ is the density of gold, $E_{\rm F} = 5.52$~eV.
For $\Delta E_{\rm el} = h\nu = 5.6{\cdot}10^{-5}$~meV, $D = 162.1$~nm.
Remind that these are estimates for the smallest GNP size which allows the RF absorption at 13.56~MHz, from the energy level spacing arguments. Now, we turn
to the momentum uncertainty arguments and assume for simplicity that the direction 
along the nanoparticle's diameter is collinear with the momentum
$\Delta p_{\rm F}$ and the Fermi electron's momentum $\vec{p}_{\rm F}$.
Then the uncertainty in the Fermi electron's momentum in this direction equals
$\Delta p \sim h/(2\pi{\cdot}D)$. At $D = 162.1$~nm, 
$\Delta p\,\sim\,6.5{\cdot}10^{-23}$~g$\cdot$cm$\cdot$s$^{-1}$, that largely exceeds
$\Delta p_{\rm F}\approx 6.4{\cdot}10^{-28}$~g$\cdot$cm$\cdot$s$^{-1}$.
Consequently, the momentum conservation law can be helped by such uncertainty,
and the RF photon absorbed. We keep in mind, however, that the minimal GNP size
for the direct RF absorption must be at least $\approx$162.1~nm. Obviously as
the particle size grows, the distance between quantified levels shrinks,
so that an absorption of a given photon $h\nu$ may bring about an excitation across
$\alpha$ (an integer number) interlevel distances. This situation is schematically 
shown in Fig.~\ref{fig_03}, which implies a parabolic dispersion law $E = p^2/2m$
for different confinements (particle sizes), in arbitrary scale. The effect of
the momentum uncertainty is indicated by dots of different widths placed at
the initial and final levels. Some numerical estimates for the energy spacings and 
momentum uncertainties, relevant for the absorption of quanta at $\nu=13.56$~MHz
($h\nu$ = 5.6$\cdot$10$^{-5}$~meV) are summarized in Table~\ref{tab:1}.
An issue of matrix elements for such transitions, some of which might be suppressed
in the dipole approximation due to symmetry considerations, is left aside.
The numerical relation between the energy split multiplicity $\alpha$,
given by $\Delta E_{\rm el} = h\nu/\alpha$, and the diameter of particle
(assuming the latter spherical) from Eq.~(\ref{eq:D-a}) is \linebreak as follows:

\begin{table}[b]
\caption{Parameters of GNPs with level spacings $h\nu/\alpha$
($\alpha$=1, 2, 4, 10 taken as examples) able to absorb the 13.56~MHz photons
thanks to the uncertainties in the Fermi electrons' momenta}
\medskip
\label{tab:1}
\begin{tabular}{p{3.2cm}p{3.8cm}p{2.6cm}p{4.1cm}}
\hline\noalign{\smallskip}
Number of \newline level spacings $\alpha$ &
Level spacing \newline $h\nu/\alpha$ ($10^{-5}$~meV) &
Diameter $D_{\alpha}$~(nm) &
Uncertainty in Fermi electron's momentum
${\Delta}p_{\alpha}$ ($10^{-23}$~g$\cdot$cm$\cdot$s$^{-1}$) \\
\noalign{\smallskip}\hline\noalign{\smallskip}
~1 & 5.6 & 162.1 & 6.5 \\
~2 & 2.8 & 204.2 & 5.2 \\
~4 & 1.4 & 257.3 & 4.1 \\
10 & 0.6 & 349.2 & 3.0 \\
\noalign{\smallskip}\hline\noalign{\smallskip}
\end{tabular}

\end{table}

%
\begin{equation}
D \approx \alpha^{1/3}
{\cdot}[(8/\pi){\cdot}(m_a/\rho){\cdot}(E_{\rm F}/h\nu)]^{1/3}
= \alpha^{1/3}{\cdot}162.1\;[\mbox{nm}].
\end{equation}
The last column of Table~\ref{tab:1} specifies the corresponding momentum uncertainty, 
$h/(2\pi{\cdot}D)$. In all the cases considered (until entering much larger particle 
sizes than those covered by Table~\ref{tab:1}), the uncertainty
in the Fermi electron's momentum $\Delta p_{\alpha}$ exceeds by far the ``adjustment'' 
$\Delta p_{\rm F}\approx 6.4{\cdot}10^{-28}$~g$\cdot$cm$\cdot$s$^{-1}$,
discussed above as a reference value to absorb the momentum mismatch
required to promote a Fermi electron onto the energy level $(E_{\rm F}+h\nu)$.
Thus, the momentum conservation law can be fulfilled, and the 13.56~MHz photons
absorbed, for GNPs larger than $\sim$162~nm. Consequently the latter, as it seems,
can apparently be
RF-heated ``directly'', i.e., without involvement of LAVMs, just by exciting
the Fermi electrons to the available energy levels above the Fermi level.
On the contrary, in GNPs smaller than $\sim$162~nm the separation between
adjacent levels exceeds the energy of the RF quantum, and at ``working'' particle
sizes which in fact show an enhanced absorption ($\sim$5~nm) the level splitting
becomes prohibitively large. An involvement of many photons in a single electron
excitation is likely to be a very rare process. Therefore, we have to bring another
physical mechanism into consideration. Longitudinal acoustic phonons seem to be able
to intervene with energies of the ``correct'' order of magnitude.
The prerequisites of their practical involvement are discussed in the following.

\section{LAVM-assisted absorption of RF photon by a Fermi electron}
\label{sec:3}
The observed size dependence of the HR can be explained by a mechanism
of the GNPs' heating that attributes a crucial role in the absorption of a RF photon
to LAVM. To make reliable estimates, we need to know something about elastic properties
of nanoparticles. A theoretical work by D.Y.~\citet{PRB63-193412}
demonstrated that the metallic nanoparticles retain the bulklike core region.
Experimentally, J.~\citet{NatMater13-1007} have recently found
that Ag nanoparticles can be deformed like liquid droplets
but remain highly crystalline in the interior.

As was already admitted above, we assume the GNP to be spherical; we remind that
the RF of 13.56~MHz corresponds to photon energy
$h\nu=5.6{\cdot}10^{-8}$~eV and momentum
$p_{\rm ph}{\approx}3{\cdot}10^{-30}$~g$\cdot$cm$\cdot$s$^{-1}$.
The absorption happens due to transitions between discrete levels
induced by quantum confinement, as was elaborated above and schematically shown
in Fig.~\ref{fig_03}. 

From now on, we turn to discussing a case of nanoparticles too small
for a RF phonon energy to ``bridge the gap'' between largely split quantified levels.
We bring into consideration a scheme of absorption of a RF photon whereby
a Fermi-level electron excitation is helped by an involvement of a LAVM.
A ``classical'' view onto the electron-phonon interaction (otherwise
straightforwardly grasped in terms of energy and momentum exchange 
between the corresponding quasiparticles) is that the electrons are driven by,
or themselves contribute to, the fluctuating electric field due to
compression / dilation of the electron density in the course of lattice vibrations.  
Obviously, only the longitudinal vibration mode can be ``useful'' in this sense.
Its wavevector moreover must be reasonably far from the Brillouin zone (BZ) center,
where the dispersion starts to bend ``flat'' and to yield high density
of modes.\footnote{%
A numerical estimate can be drawn from \protect\cite{IJAPM3-275},
who studied the phonon dispersions in amorphous metals. For gold, the first maxiumum
of the $\omega(q)$ dispersion occurs at $q{\sim}$1.5~{\AA}$^{-1}$ whereas
up to $q{\sim}$1.0~{\AA}$^{-1}$ the dispersion remains reasonably linear.}

\begin{figure}[t]
\parbox[b]{9.0cm}{\includegraphics[width=8.2cm]{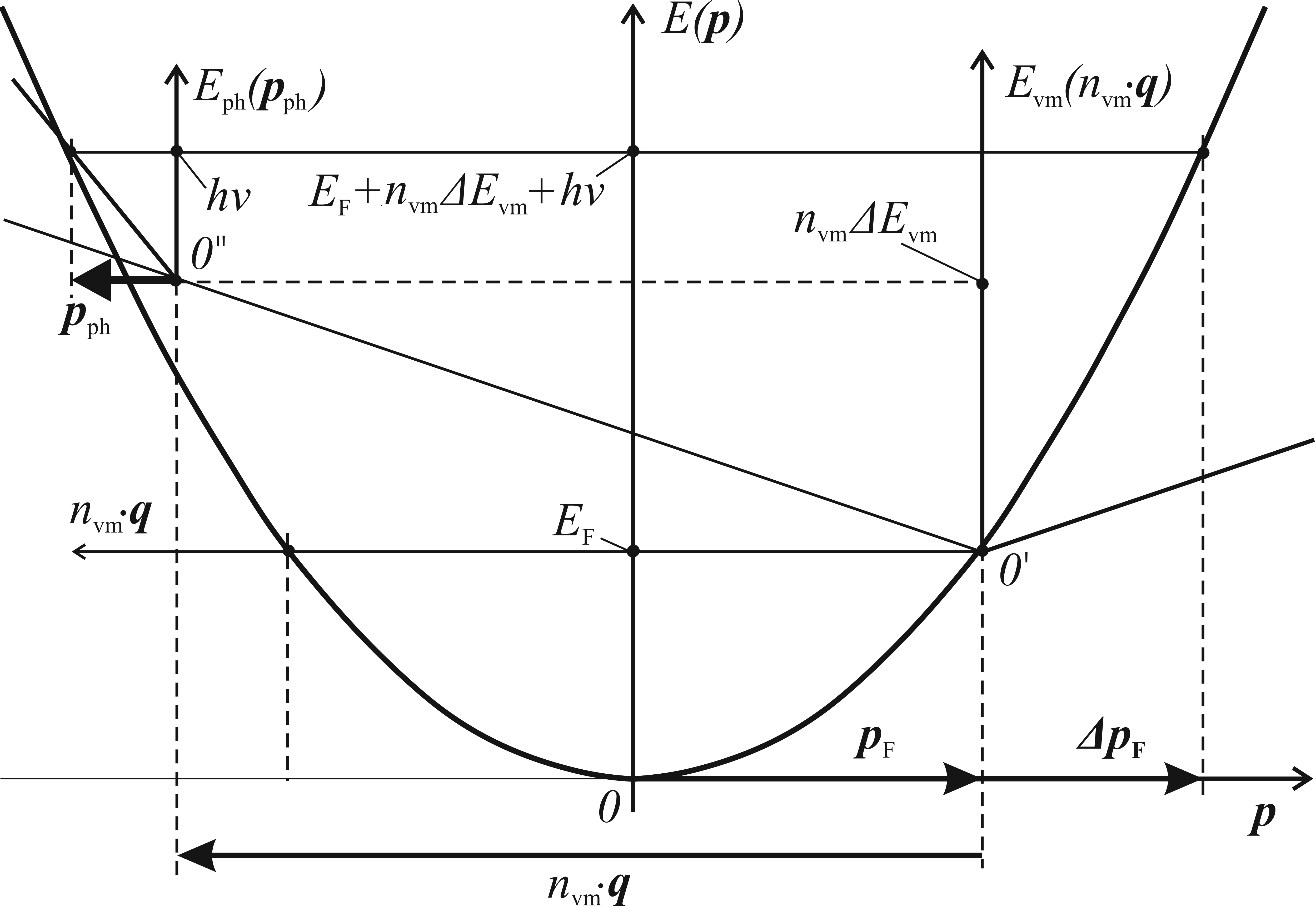}}
\parbox[b]{7.0cm}{
\caption{Scheme of absorption of RF photon by the Fermi electron
assisted by an absorption of a LAVM (in arbitrary scale). The dispersion laws
for photons $E_{\rm ph}{\sim}p_{\rm ph}$, electrons $E(p){\sim}p^2/2m$
and LAVMs $E_{\rm vm}{\sim}(n_{\rm vm}{\cdot}q)$ are shown. See text for details.} 
\label{fig_04}}       
\end{figure}

A Fermi electron may absorb the energies of both
the RF photon and the LAVM. Fig.~\ref{fig_04} shows how an inclusion of momentum
$n_{\rm vm}{\cdot}q$ and energy $n_{\rm vm}{\Delta}E_{\rm vm}$ of such LAVM
($n_{\rm vm}$ numbers the energy steps in the vibration modes)
into the combined absorption makes possible to satisfy both conservation laws. 
In the absorption event, the GNP borrows energy
from its LAVMs system, adds it to that of the RF quantum and excites
an electron to the vacant level beyond the $E_{\rm F}$:
\begin{equation}
E_{\rm F}+\Delta E_{\rm vm}+h\nu = (\vec{p}_{\rm F}+\vec{q}+
\vec{p}_{\rm ph})^2/2m\,,
\label{eq:01}
\end{equation}
where $E_{\rm F} / \vec{p}_{\rm F}$, $\Delta E_{\rm vm} / \vec{q}$
and $h\nu / \vec{p}_{\rm ph}$ are the energy / momentum of the free electron,
the LAVM and photon, correspondingly (see Fig.~\ref{fig_04}),
and $m$ is the electron's mass. 
We note, for the sake of a later reference, that an electron may well be
excited across $m_{\rm el}>1$ steps of its discrete spectrum, i.e.,
onto $E_{\rm F}+m_{\rm el}\,{\Delta}E_{\rm el}$, that is however not explicitly
depicted in Fig.~\ref{fig_04}. Once excited, the propagating mobile electron  
very likely will be trapped (relaxed) before reaching the GNP surface
(see Subsec.~\ref{subsec:free_path} on the mean free path issues),
releasing more energy to the LAVMs pool than was borrowed beforehand.
The net effect of that is the heating of the GNP.

In Fig.~\ref{fig_04}, a point on the Fermi sphere (identified by momentum
$\vec{p}_{\rm F}$ and energy $E_{\rm F}$) serves as an origin of
the LAVM dispersion branch, on which the momentum $n_{\rm vm}\!\cdot\!\vec{q}$
and the mode energy
${\Delta}E_{\rm vm}$ stand for some ``representative'' mode. 
Let us give some numerical estimates to these parameters.

The free-electron model for bulk gold yields the Fermi radius 
$k_{\rm F}=1.20{\cdot}10^{10}$~m$^{-1}$, quite close to the mean experimental
estimates for its slightly non-spherical belly \citep{PRB25-7818}. The corresponding 
free-electron Fermi energy is $E_{\rm F}=5.52$~eV; the Kubo formula for the level
splitting at the Fermi energy due to spatial confinement 
${\Delta}E_{\rm vm}$  yields, for the diameter $D=5$~nm,
${\Delta}E_{\rm vm}=1.91$~meV, and the corresponding increment
of the electron's momentum to bring it to the first vacant level is
${\Delta}k_{\rm F}=2.08{\cdot}10^6$~m$^{-1}\,\approx\,0.017$ of the $k_{\rm F}$.
As RF $h{\nu}=5.6{\cdot}10^{-5}$~meV$\ll{\Delta}E_{\rm vm}$, the larger part
of the latter gap has to be overcome, according to out hypothesis,
by borrowing energy from LAVM. From the  velocity of sound in gold
(3240~m$\cdot$s$^{-1}$ longitudinal, 1200~m$\cdot$s$^{-1}$ transversal), 
the highest phonon energy extrapolated onto the nearest BZ boundary, i.e.,
the $L$ point, is ${\sim}28.5$~meV ($L_{\rm L}$) and ${\sim}10.5$~meV ($L_{\rm T}$).
Corresponding experimental frequencies at $L$ are bended downwards as expected,
to, correspondingly, 19.1 and 7.7~meV 
\citep[as cited by][see also his first-principles calculation
of vibration spectrum]{JPCM25-145401}. 
Consequently, the energy matching to
${\Delta}E_{\rm vm}$ can be realized via an interaction with an acoustic phonon
whose momentum is within ${\sim}7-18$\% of the BZ radius. Inversely,
an interaction with a phonon close to the BZ periphery may, in principle, promote
an electron onto a higher (up to the 3th or 4th) energy state beyond the Fermi level.
Two observations, however, need to be stressed in this relation. First,
the numerical relations in gold (see above) are such that matching the energies
of electron and phonon is impossible if their momenta stay collinear,
as Fig.~\ref{fig_04} implies. A more realistic scenario, discussed below,
comprises $\vec{k}_{\rm F}$ staying at some angle to the phonon momentum $\vec{q}$.
The second observation imposes the quantisation of phonon $\vec{q}$ values
in a nanoparticle as $q = h/L$, where $L$ is a length of the chain of ions
along which the LAVM propagates. As an estimate of the order of magnitude,
the relevant length is between the nanopaticle's diameter (5~nm) and circumference,
hence the $q$ step is 
${\Delta}q=[1{\cdots}2\pi]/L=[0.2{\cdots}1.26]{\cdot}10^9$~m$^{-1}$,
i.e., ${\sim}[2{\cdot}11]$\% of $k_{\rm F}$. 

In the following, we'll often refer to circular contour / path over which
the compression mode propagates, as this is a simple yet realistic model case.
Although the natural diversity of nanoparticle sizes and shapes makes a faithful
simulation difficult, the circular path has a virtue of being the longest one
in a ``round'' particle, hence hosting the maximal number of modes,
densely distributed in the $\vec{q}$ space. Consequently, the quantified energies
of modes on a circular contour make a denser spectrum than those on any other path; 
within a given energy interval, more individual modes can be found and used
for borrowing energy to an electron. It will be argued below that longitudinal
acoustic phonon cannot propagate strictly on the surface, but rather
at a (small) depth. 

\begin{figure}[t]
\parbox[b]{8.8cm}{\includegraphics[width=7.2cm]{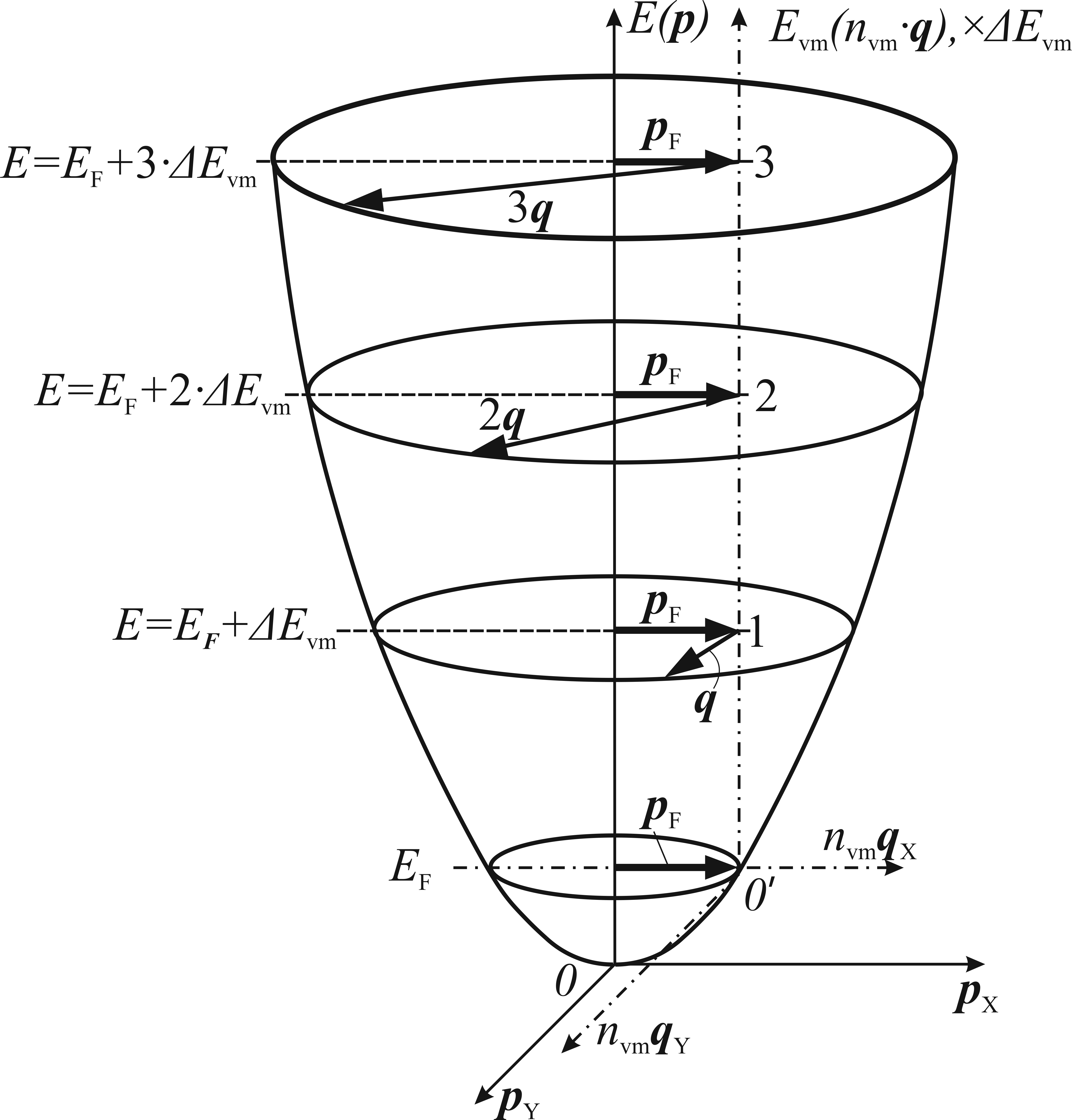}}
\parbox[b]{7.2cm}{%
\caption{Same as in Fig.~\ref{fig_04}, 
however assuming $m_{\rm el}=n_{\rm vm}$ and
${\Delta}E_{\rm el}={\Delta}E_{\rm vm}$,
for the general case of the electron momentum
$\vec{p}_{\rm F}$ and the phonon momentum $\vec{q}$ being coplanar. For simplicity,
the energy and the momentum of the RF photon are neglected.
See text for details.}
\label{fig_05}   }    
\end{figure}

We discuss now the impact of the electron and phonon momenta being non-collinear,
but, in any case, coplanar. A more realistic scheme in this sense than that
of Fig.~\ref{fig_04} is shown in Fig.~\ref{fig_05}, where however the energy
and the momentum of the RF photon are neglected, by force of relations
$h{\nu}\,\ll\,E_{\rm vm}$; $p_{\rm ph}\,\ll\,n_{\rm vm}\,q$; 
$p_{\rm ph}\,\ll\,p_{\rm F}$. Fig.~\ref{fig_05} depicts therefore 
a possibility of absorption, by electrons at the Fermi surface, of the LAVMs with
energies ${\Delta}E_{\rm vm}$, $2{\Delta}E_{\rm vm}$, $3{\Delta}E_{\rm vm}$,
and corresponding momenta, within the same nanoparticle of a given size.
It is (arbitrarily, just for the sake of simplifying the figure) implied that
$m_{\rm el}=n_{\rm vm}$ and hence ${\Delta}E_{\rm el}={\Delta}E_{\rm vm}$,
therefore the augmentation of the electron energy on absorption happens in portions
of ${\Delta}E_{\rm vm}$.

\begin{figure}[b]
\includegraphics[width=12.0cm]{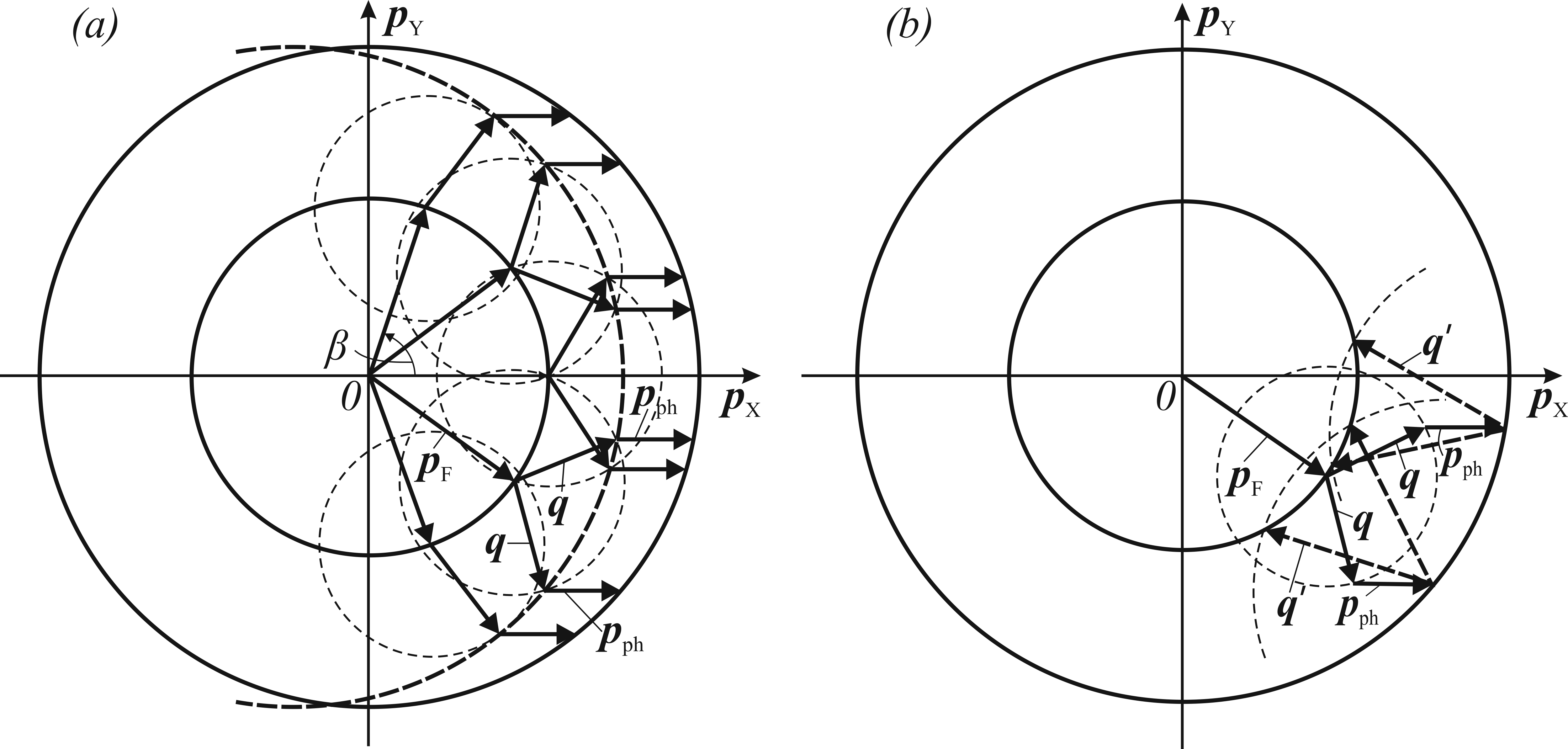}
\caption{Projection of the vector construction from Fig.~\ref{fig_05}
onto the plane comprising the $\vec{p}_{\rm F}$ and $\vec{q}$ vectors;
the photon momentum $\vec{p}_{\rm ph}$, neglected in Fig.~\ref{fig_05},
is now retained. Circles centered at the origin of the coordinate system,
of radii $|\vec{p}_{\rm F}|$ and $|\vec{p}_{\rm F}+\vec{q}+\vec{p}_{\rm ph}|$,
are cross-sections of the electron energy paraboloid at $E=E_{\rm F}$ and
$E=E_{\rm F}+{\Delta}E_{\rm vm}+h\nu$. ($a$): Some examples of matching
momenta and energies in the process of electron excitation from $\vec{p}_{\rm F}$
due to absorption of phonon $\vec{q}$ and RF photon $\vec{p}_{\rm ph}$.
($b$): For one of $\vec{p}_{\rm F}$ vectors from panel ($a$),
excitation channels are followed by possible relaxation channels into the initial
state via release of phonon $\vec{q}'$. 
Note that $|\vec{q}'|>|\vec{q}|$.}
\label{fig_06}       
\end{figure}

A more detailed projection of the momenta matching,
involving also the RF photon momentum, is shown in Fig.~\ref{fig_06}.
Possible excitations from an initial state $\vec{p}_{\rm F}$ of energy
$E_{\rm F}$ via an absorption of a phonon $\vec{q}$ and a RF photon
$\vec{p}_{\rm ph}$ end up in a state with the energy 
$E = E_{\rm F}+{\Delta}E_{\rm vm}+h{\nu}$. The allowed ``chained''
$\vec{p}_{\rm F}{\cdots}\vec{q}{\cdots}\vec{p}_{\rm ph}$ vectors
fall within a body of revolution around the fixed 
$\vec{p}_{\rm ph}{\parallel}\vec{p}_x$ direction, limited on the left by the cone
of [side $\vec{p}_{\rm F}$ / aperture $2\beta$] and on the right --
by the spherical (radius $|\vec{p}_{\rm F}+\vec{q}+\vec{p}_{\rm ph}|$) cape
built on top of the cylinder of the height $p_{\rm ph}$. The conical
and cylindrical surface parts are connected by the intermediate conical belt
of the width $q$.  From Fig.~\ref{fig_06}(a), the maximal angle
$\beta$ the $\vec{p}_{\rm F}$ vector may build to $\vec{q}$ is

\vspace*{-4mm}
\begin{equation}
\beta=\arccos\frac{(\sqrt{2m_eE}-q)^2-p_{\rm ph}^2-p_{\rm F}^2}
{2(\vec{p}_{\rm ph}{\cdot}\vec{p}_{\rm F})}\,.
\end{equation}
On relaxation of the excited electron back to $E_{\rm F}$, the extra energy
$E'={\Delta}E_{\rm vm}+h\nu$ is released into the vibration pool,
exciting a phonon with $|\vec{q}'|>|\vec{q}|$. $h\nu$ is therefore
the net gain in energy.

\begin{figure}[t]
\parbox[b]{8.8cm}{\includegraphics[width=8.0cm]{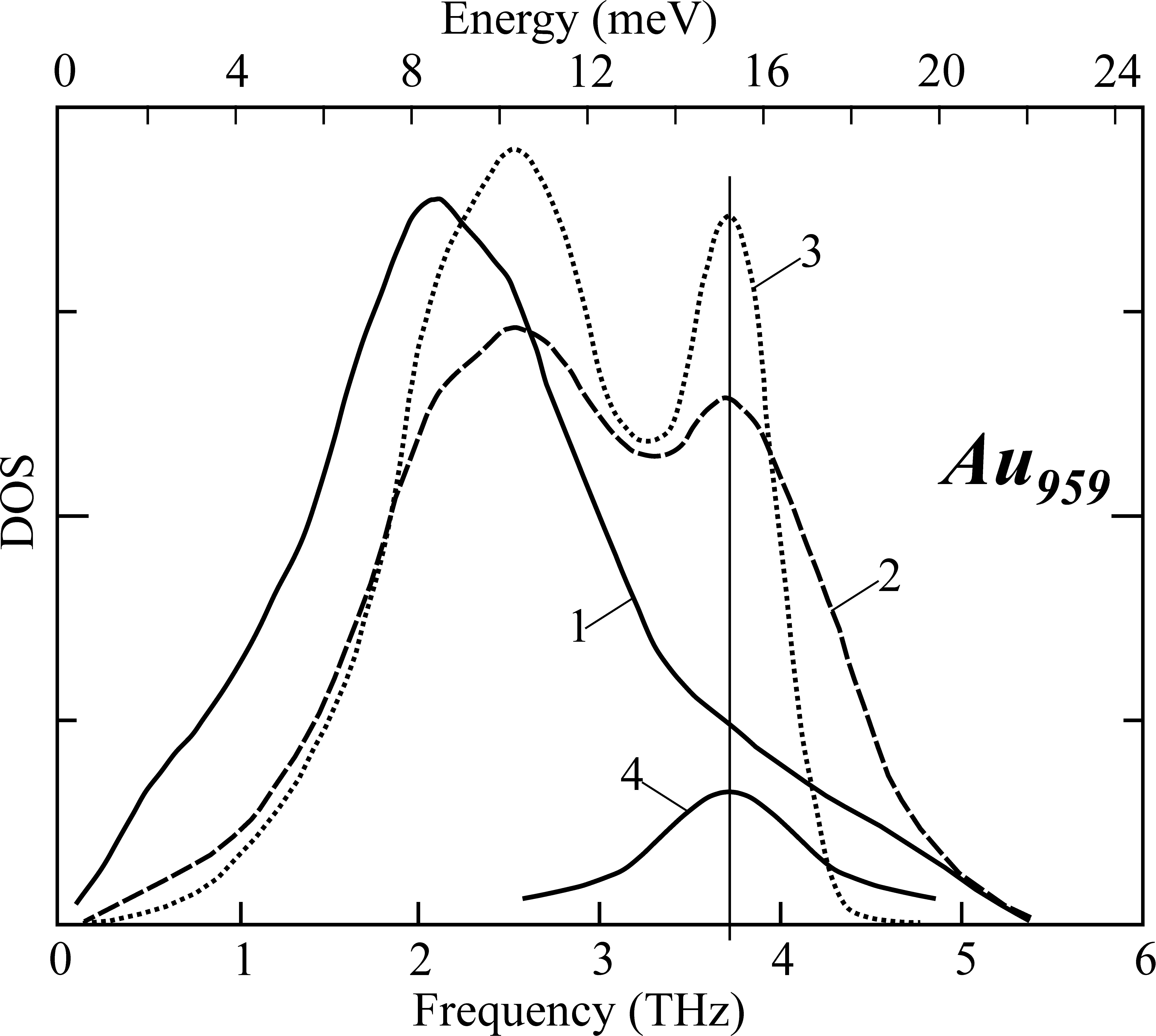}}
\parbox[b]{6.5cm}{%
\caption{VDOS in Au$_{959}$ nanocrystal, reproduced with our modification
with permission from \cite{PRB63-193412}. 
Curve 1: contribution from the surface shell;
curve 2: from the transition shell, curve 3: from the core region; curve 4:
our approximate extraction of the VDOS for LAVMs in the core region, yielding a peak
centered at 3.7~THz (15.3~meV), with full width at half maximum (FWHM)
$\sim$1~THz ($\sim$4.14~meV).}
\label{fig_07}}    
\end{figure}

\begin{figure}[b]
\parbox[b]{8.8cm}{\includegraphics[width=8.0cm]{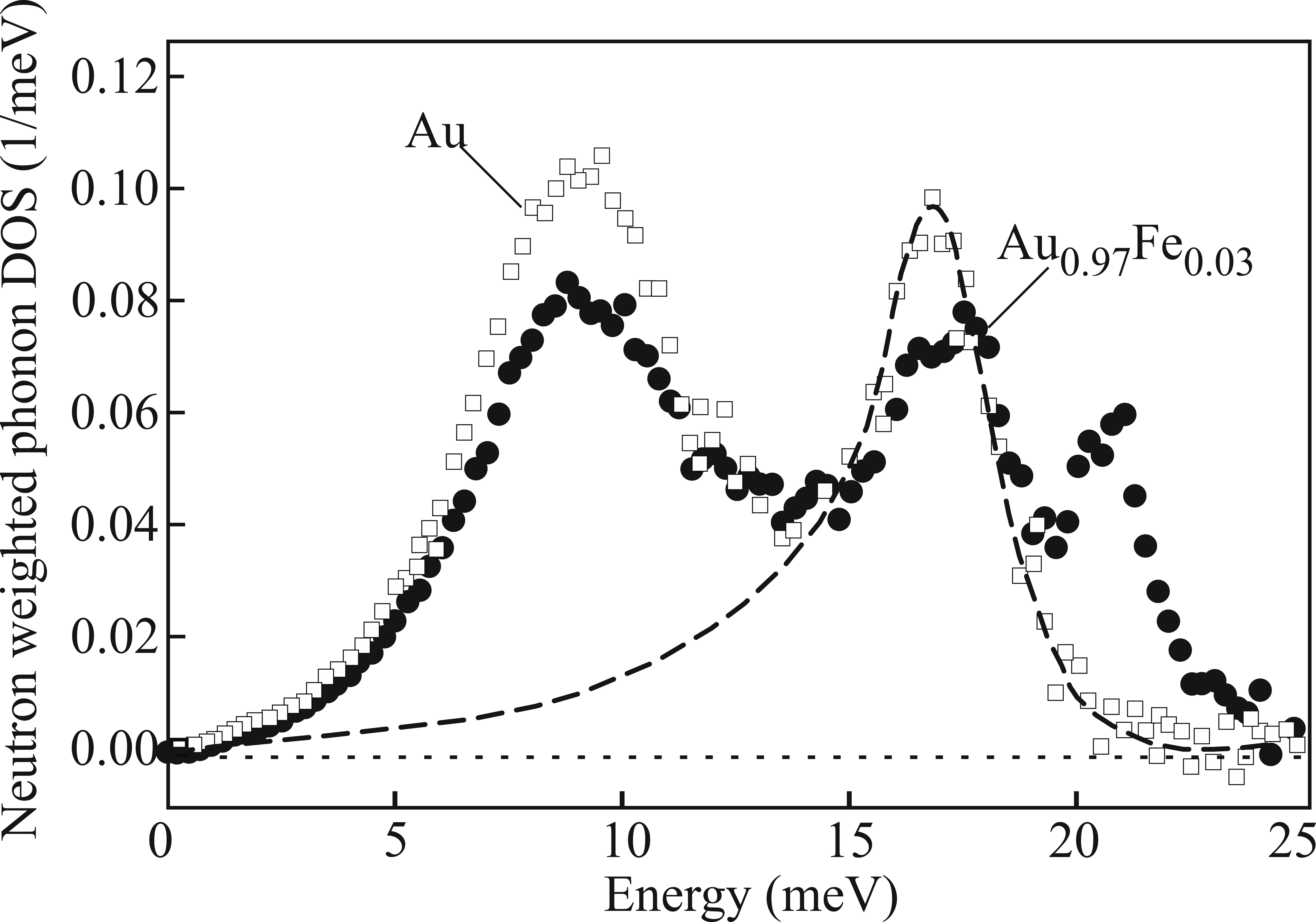}}
\parbox[b]{6.5cm}{%
\caption{Experimental phonon DOS curves for bulk pure Au and Au$_{0.97}$Fe$_{0.03}$
from the inelastic neutron scattering measurements; reproduced with permission
from \cite{PRB87-014301}. Dashed line is our approximate extraction of the VDOS curve 
for LAVMs (the FWHM range is from 14.6 to 18.4~meV; the peak position is at 16.8~meV).}
\label{fig_08}} 
\end{figure}

In the following, we consider the LAVM within a GNP as a compression wave
propagating along a closed chain of atoms, in the spirit of 
the Born -- von K{\'a}rm{\'a}n cyclic boundary conditions. 
For this analysis, the work by D.Y.~\cite{PRB63-193412}
is useful which deals with elastic properties and
vibrational density of states (VDOS)
in GNPs, indicating notably three structure elements of a nanoparticle:
the surface shell (of $\approx\,2$~{\AA} thickness), the transition shell
(of $\approx\,3$~{\AA} thickness beneath) and the core region.
Fig.~\ref{fig_07} depicts the corresponding decomposition of VDOS
for the case of 959-atom GNP, of the diameter $\approx\,3.2$~nm. In the core and 
transition shell, the distribution of VDOS clearly shows the peaks at
$\approx\,3.7$~THz ($E_{\rm vm}\,\approx\,15.3$~meV), resembling the peak in the VDOS
of bulk gold in the experimental work by \cite{PRB87-014301},
which we attribute as being due to LAVMs
(see Fig.~\ref{fig_08}). The VDOS of the surface shell in Fig.~\ref{fig_07} does also
indicate a feature at the corresponding energy, that we extracted
from the background as the curve 4 in Fig.~\ref{fig_07}.

The VDOS maximum under discussion is strongest in the core region, which is therefore 
expected to contribute mostly to the RF absorption. According to 
D.Y.~\cite{PRB63-193412}, the core region is bulklike,
therefore while discussing its contribution to the RF absorption, one can rely
on the density and sound velocity values for crystalline gold. This is consistent
with experimental observations on the gold's homologue, silver,
by J.~\cite{NatMater13-1007}
who have found that its nanoparticles remain highly crystalline in the interior. 
In the following discussion, we assume that the ``useful'' LAVMs propagate
at the depth ${\delta}=0.7$~nm, i.e., along the closed contour entirely
within the bulklike core region. 

\subsection{Energy conservation}
We discuss now some quantization relations for electrons and LAVMs following
from the GNP geometry. The energy conservation condition for the absorption
of a RF photon with the energy $h\nu$ is as follows:
\begin{equation}
m_{\rm el}{\cdot}{\Delta}E_{\rm el} = n_{\rm vm}{\cdot}{\Delta}E_{\rm vm}
+h\nu \approx n_{\rm vm}{\cdot}{\Delta}E_{\rm vm}\,. 
\label{eq:mEl}
\end{equation}
Here, $m_{\rm el}$ is the number of steps (gaps) 
between the quantized electron levels,
and $n_{\rm vm}$ the number of vibration quanta helping a RF photon to get absorbed. 
According to the Kubo's formula \citep{JPSJ17-975,JPhysColloq38-C2-69} --
see also Eq.~(\ref{eq:A4}) in the Appendix \ref{sec:App1}, -- 
the step in the electron energy
levels ${\Delta}E_{\rm el}$ depends on the number of gold atoms $N_a$.
The condition (Eq.~\ref{eq:mEl}) that $m_{\rm el}$ energy steps must embrace
$n_{\rm vm}$ phonon energies takes the form
\begin{equation}
\frac{4}{3}\,m_{\rm el}\,\frac{E_{\rm F}}{N_a}\approx 
\frac{n_{\rm vm}\,v_{\rm L}\,h}{\pi(D-2\delta)}\,,
\label{eq:m_el1}
\end{equation}
where $L=\pi(D-2\delta)$ is the length of the closed contour at the depth $\delta$
under the GNP surface, and the (longitudinal) sound velocity $v_{\rm L}$ 
relates frequency to wave vector. Further on, assuming that the density of gold
in the surface shell  and the transition shell is close to that in the core region, 
i.e. in bulk gold, the number of atoms in GNP can be expressed via density of gold
$\rho$, atomic mass $m_a$ and the volume $V$ (or diameter $D$) of the particle:
\begin{equation}
N_a\,{\approx}\,\frac{\rho V}{m_a}=\frac{\pi}{6}\,\frac{\rho}{m_a}D^3\,.
\end{equation}
Taken together with Eq.~(\ref{eq:m_el1}), this yields the depressed
cubic equation on $D$:
\begin{equation}
D^3+\mathfrak{p}\,D+\mathfrak{q}=0\,,
\label{eq:D3pD}
\end{equation}
\begin{equation}
\mbox{with}\quad
\mathfrak{p}
\,=-\frac{8E_{\rm F}\,{m_a}}{v_{\rm L}\,h\rho}
\!\left(\!\frac{m_{\rm el}}{n_{\rm vm}}\right)\,;
\quad\quad
\mathfrak{q}
\,=\frac{16E_{\rm F}\,{m_a}\delta}{v_{\rm L}\,h\rho}
\!\left(\!\frac{m_{\rm el}}{n_{\rm vm}}\right)\,.
\end{equation}
%
For $\mathfrak{p}<0$ and $\mathfrak{q}>0$,
Eq.~(\ref{eq:D3pD}) always has one negative root and either two
(possibly degenerate) or none positive ones. The two positive roots
of Eq.~(\ref{eq:D3pD}) can be expressed as follows:
\begin{equation}
D_k = -2\sqrt{\frac{\bigl|
\mathfrak{p}
\bigr|}{3}}
\cos\Bigl\{\frac{1}{3}\arccos\Biggl[\,
\frac{
\mathfrak{q}
}{2}\!\left(\!\frac{3}{\bigl|
\mathfrak{p}
\bigr|}
\!\right)^{\mbox{\hspace*{-3pt}}\frac{3}{2}}\,\Biggr]+\frac{2\pi}{3}k\Bigr\}\,,
\end{equation}
where $k=1,2$. Note that the parameters $\mathfrak{p}$ and $\mathfrak{q}$
contain, along with constants depending on the properties of gold,
the trial numbers $\delta$, $m_{\rm el}$ and $n_{\rm vm}$;
recall that $\delta$ is the depth of propagation of LAVM, and $n_{\rm vm}$
is the number of vibration quanta matching within the electronic excitation.
The solutions  $D_k$ can be expressed in terms of $\delta$, $m_{\rm el}$
and $n_{\rm vm}$ as follows ($D_k$ and $\delta$  measured in nm):

\begin{figure}[t]
\parbox[b]{8.0cm}{\includegraphics[width=7.2cm]{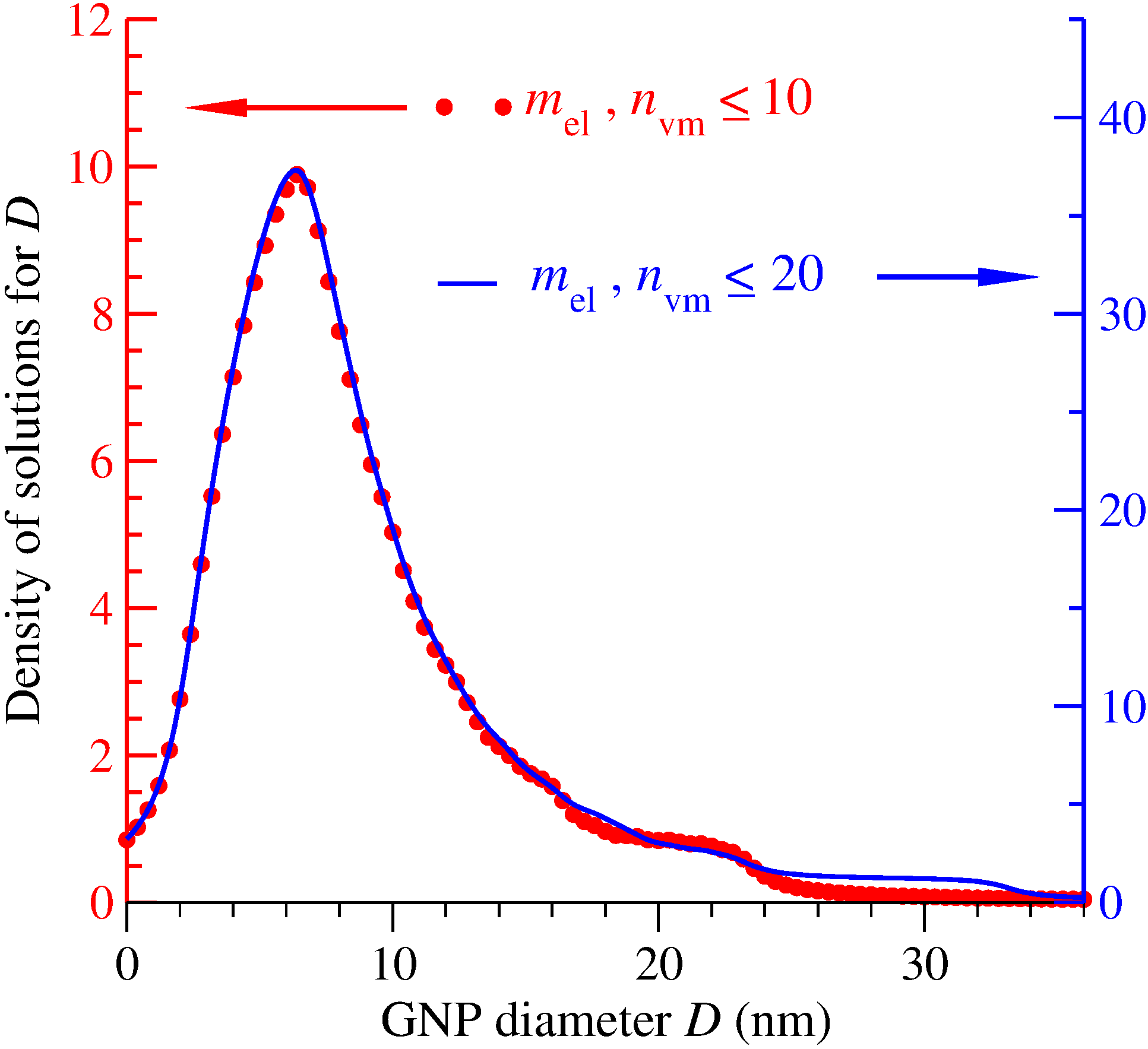}}
\parbox[b]{7.0cm}{
\caption{%
The density of discrete solutions $D_{k=1}$ after Eq.~(\ref{eq:Dk_num}),
smeared into a continuous distribution with the half-width parameter of 1~nm,
for better visibility.
$\delta$ is fixed to 0.7~nm.
The cases $(m_{\rm el}, n_{\rm vm}){\leq}10$ (red dots, refer to the left
scale) and $(m_{\rm el}, n_{\rm vm}){\leq}20$
(blue continuous line, refer to the right scale) are compared.
See text for details.}
\label{fig_09}}
\end{figure}

%
\begin{equation}
D_k = -8.64\sqrt{\frac{m_{\rm el}}{n_{\rm vm}}}\cos\Bigl[\frac{1}{3}\arccos\Bigl(
0.695{\cdot}{\delta}\sqrt{\frac{n_{\rm vm}}{m_{\rm el}}}\,\,\Bigr)
+\frac{2\pi\,k}{3}\Bigr]\,.
\label{eq:Dk_num}
\end{equation}
\noindent
Of two positive roots given by Eq.~(\ref{eq:Dk_num}), we retain
the practically relevant largest value of $D_k$, corresponding to $k=1$.
Note that certain combinations ($m_{\rm el}$, $n_{\rm vm}$) yield
the argument of arccosine $>1$ and hence no solution. The roots are densely
yet unevenly distributed, as is shown by Fig.~\ref{fig_09}. Note that
the total number of solutions increases, as expected, with the number
of trial ($m_{\rm el}$, $n_{\rm vm}$) combinations, however the profile
of the root density, with its narrow maximum around $\approx\,7$~nm,
remains stable up to the upper $\sim\,25$\% of the total span of $D$ values.
For this reason, and moreover since a non-ideal equidistanteness
of both electrons' and LAVMs' levels would eventually ``detune'' the criteria 
set by Eqs.~(\ref{eq:mEl}, \ref{eq:m_el1}) for large ($m_{\rm el}, n_{\rm vm}$),
we set, from now on, an arbitrary limit ($=10$) on the maximal values of
the latter.

We'll see below that this remarkable ``selectivity'' of GNP sizes $D$
with respect to their capacity to satisfy the energy conservation relations
will eventually manifest itself in the distribution of the HR. 
The latter ought to be influenced, however, by the next element
entering our discussion, namely, the availability of ``useful'' LAVMs
that can be induced in the particles of selected sizes. 
``Useful'' means the modes whose energies are multiples of  
$n_{\rm vm}{\cdot}{\Delta}E_{\rm vm}=n_{\rm vm}{\cdot}v_{\rm L}h/[\pi(D-2\delta)]$,
$n_{\rm vm}$ being selected by the commensurability of the vibration
energies with electron excitation ones, in the spirit of Eq.~(\ref{eq:mEl}).
Obviously, for the optimal heating it is essential to find many such vibrational modes
within the FWHM of the LAVM-related VDOS peak of gold
(cf. Figs.~\ref{fig_07},\ref{fig_08} and the related discussion),
the number we'll refer to as $N_{\rm FWHM}$ in the following.

\subsection{Momentum conservation}
\begin{figure}[b]
\parbox[b]{7.6cm}{\includegraphics[width=6.6cm]{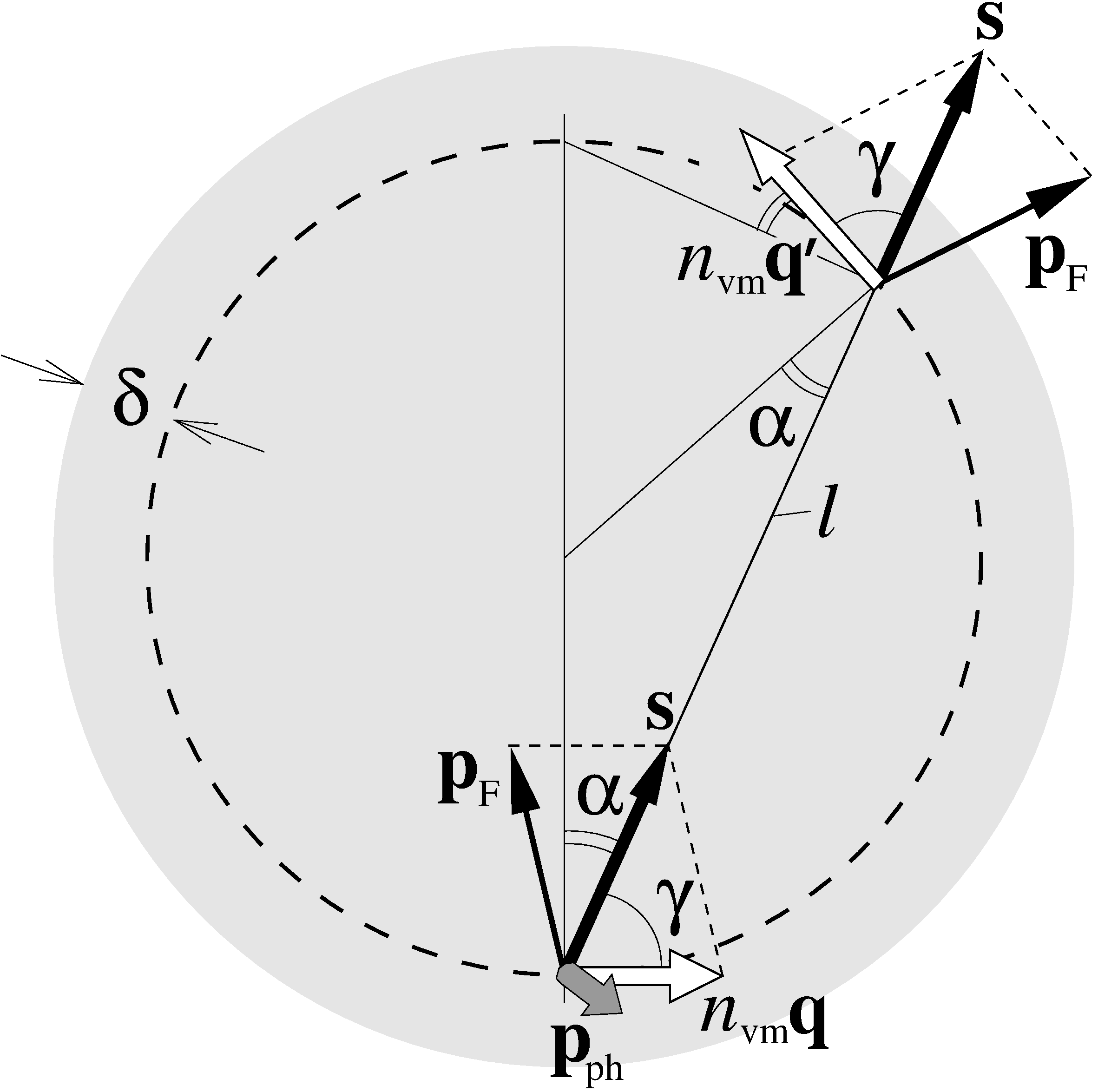}}
\parbox[b]{8.0cm}{
\caption{Scheme of RF photon absorption on a circular contour inside a nanoparticle; see text for discussion.}
\label{fig_10}}   
\end{figure}

In addition to the energy conservation equation (\ref{eq:mEl}), one should
take into account the momentum conservation condition. We'll specify it
for the case of LAVM propagating along the circular contour of the diameter
$D-2\delta$, to which the phonon momentum $n_{\rm vm}\vec{q}$
is tangential -- see Fig.~\ref{fig_10}. Anywhere on the contour, a Fermi electron
with the momentum $\vec{p}_{\rm F}$ can intervene to bring about an absorption
of a RF photon (we neglect the effect of the momentum and energy of the latter
onto the resulting conservation relation, as was already argued before). 

The situation arbitrarily shown in Fig.~\ref{fig_10} assumes the momentum
of the Fermi electron to point inwards the GNP; on absorbing a phonon,
such electron would transverse the particle along the path $l$ and release
the phonon ``on the other side'' of the contour. A different possibility
would be the $\vec{p}_{\rm F}$ pointing outwards; on the nanoparticle's
surface such electron
would be either elastically reflected inwards, or emit a quantum
and be ``lost'' for the GNP heating mechanism we discuss. The probability
relation for such elastic / inelastic events at the surface is \emph{a priori}
difficult to estimate. Anyway, the elastically scattered electron will almost
``instantly'' regain the contour and follow the path of the inwards-moving electron,
as depicted in Fig.~\ref{fig_10}. To specify what ``instantly'' means,
we note that the perturbation of the potential inside the particle,
``felt'' by an electron, propagates, at most, with the longitudinal
sound velocity in gold, i.e., $v_{\rm L}=3.23{\cdot}10^5$~cm$\cdot$s$^{-1}$,
whereas the Fermi velocity in gold is three orders of magnitude larger:
the free-electron model with the electron density corresponding to that of
monovalent gold yields $v_{\rm F}=1.4{\cdot}10^8$~cm$\cdot$s$^{-1}$
\citep[cf. Sec.~\ref{sec:2}, see also][]{AshMerm_book}. 
The ``flight time'' for the electron to reach the surface of the nanoparticle
and get back to the contour is therefore ${\sim}10^{-15}$~s. 

A simple geometric argument illustrated by Fig.~\ref{fig_10}
(for the ideal case of planar circular contour) shows that the momentum
of the absorbed phonon $n_{\rm vm}\vec{q}$ can ``ride''
the electron across the particle and be released on the other side of the contour,
that we'll refer to as relaxation.
In fact, simultaneously released are the Fermi momentum $\vec{p}_{\rm F}$
and the phonon momentum $n_{\rm vm}\vec{q}'$,
both at angles with their respective ``pre-absorption'' values
but exactly preserving the corresponding moduli, under the condition
that the phonon is emitted along the contour at the electron's contact
with the latter on arrival. Specifically in Fig.~\ref{fig_10},
the transferred momentum $\vec{s}$, a sum of the electron $\vec{p}_{\rm F}$
and phonon $n_{\rm vm}\vec{q}$ momenta are related to the electron
emission angle $\gamma$ as follows:
\begin{equation}
p_{\rm F}^2 = s^2 + (n_{\rm vm}{\cdot}q)^2-2s(n_{\rm vm}{\cdot}q)\cos\gamma\,,
\end{equation}
i.e., the angle the momentum of the excited electron makes to the contour is
\begin{equation}
\gamma =\arccos
\frac{s^2+(n_{\rm vm}{\cdot}q)^2-p_{\rm F}^2}{2s(n_{\rm vm}{\cdot}q)}\,.
\label{eq:gamma1}
\end{equation}
Fig.~\ref{fig_10} implies moreover that the momentum of a LAVM phonon, 
via the interaction with an electron, may be ``reinforced''
by that of RF photon.
The ``parallel'' (along the contour) component of the latter is then added
to $n_{\rm vm}{\cdot}\vec{q}$ and transferred (by mediation of the excited electron)
to the relaxation point. The momentum of then released phonon is
$n_{\rm vm}{\cdot}\vec{q}'$, with $|\vec{q}'|>|\vec{q}|$, even if
the net increase of the phonon momentum is quite small, with respect to
electron and phonon counterparts: $h{\nu}\,\ll\,{\Delta}E_{\rm vm}$;
$p_{\rm ph}\,\ll\,n_{\rm}q$, $p_{\rm ph}\,\ll\,p_{\rm F}$. 

The magnitude of $s$ in Eq.~(\ref{eq:gamma1}) is 
$s=\sqrt{2m(E_{\rm F}+n_{\rm vm}{\Delta}E_{\rm vm})}$.
On expressing the magnitude of the LAVM momentum $q$ and the energy step
via the nanoparticle size $D$ and the velocity of (longitudinal) 
sound $v_{\rm L}$,
\begin{equation}
q= h/[\pi(D-2\delta)]\,;\quad\quad
{\Delta}E_{\rm vm}=v_{\rm L}h/[\pi(D-2\delta)]\,,
\end{equation}
we arrive at an expression for $\gamma$ in terms of $n_{\rm vm}$ and $(D-2\delta)$:
\begin{equation}
\gamma=\arccos\left\{
\left[m\,v_{\rm L}+\frac{n_{\rm vm}\,h}{2\pi(D-2\delta)}\right]\!
\left[p_{\rm F}^2+\frac{2m\,n_{\rm vm}\,v_{\rm L}h}{\pi(D-2\delta)}
\right]^{-\frac{1}{2}}\right\}\,.
\end{equation}
An excited electron would traverse the nanoparticle (along the chord $l$
in Fig.~\ref{fig_10}) and release energy on entering the cyclic contour again,
whereby a vibration mode with momentum $n_{\rm vm}\vec{q}'$ would be induced.
Technically this might happen as a consequence of an electric field being
suddenly created at the ``arrival point'' of the electron on the contour;
the Coulombic attraction of the ions would trigger the compression wave
to run along the contour. The electron path (chord) is related to
the nanoparticle parameters $n_{\rm vm}$, $D$, $\delta$ as follows: 
\begin{eqnarray}
l&=&(D-2\delta)\cos\Bigl(\frac{\pi}{2}-\gamma\,\Bigr) \nonumber \\
&=&(D-2\delta)\sin\,
\arccos\left\{\!
\left[m\,v_{\rm L}+\frac{n_{\rm vm}\,h}{2\pi(D-2\delta)}\right]\!\!
\left[p_{\rm F}^2+\frac{2m\,n_{\rm vm}\,v_{\rm L}h}{\pi(D-2\delta)}
\right]^{-\frac{1}{2}}\right\}.
\end{eqnarray}
For relevant values of $n_{\rm vm}$ and $(D-2\delta)$,
the argument of arccosine varies within ${\sim}\,0.006-0.187$.
Consequently the sine in the above formula stays within 
$0.982-0.999$, i.e., $l\,\approx\,(D-2\delta)$ with, at most, 
$\approx\,2\%$ accuracy, therefore the excited electron must transverse
the nanoparticle roughly along the latter's axis. On neglecting the discreteness
on the circle of LAVM propagation (of the $D-2\delta$ diameter), each its point
may serve as the ``source'' of the excited electron. 
The regions of absorption of RF photons in the nanoparticle are the rings 
of $(D-2\delta)$ diameter. 

\subsection{Electron free path}
\label{subsec:free_path}
The probability for an electron to transverse the particle,
i.e., to escape being scattered along the path of the length $l$ and to reach
the ``opposite'' point on the cyclic contour, equals 
$\exp(-l/l_0){\approx}\exp[-(D-2\delta)/l_0]$, where $l_0$ is
the mean free path of an electron in the nanoparticle of the size $D$
(see below).

It seems logical that the HR of a nanoparticle of size $D$ is proportional
to the following factors: 
($i$) the energy of absorbed RF photons, 
($ii$) the length of the contour  $\pi(D-2\delta)$
(as the absorption may occur in any point thereupon); 
($iii$) the summary number $\sum_i N^{(i)}_{\rm FWHM}$
of matching opportunities for the multiple
energy step $n_{\rm vm}{\cdot}{\Delta}E_{\rm vm}$ to fall within
the energy range of FWHM vibration modes [the estimates are 
$14.60-18.40$~meV according to \cite{PRB87-014301}, or $12.86-16.54$~meV
according to D.Y.~\cite{PRB63-193412}],\footnote{%
The sum $\sum_i N^{(i)}_{\rm FWHM}$ takes into consideration
contributions from the absorption of RF photons at phononic momenta
$\mathbf{q}$, $2\mathbf{q}$, $i\mathbf{q}$ etc., in the spirit of Fig.~\ref{fig_05},
for fixed $m_{\rm el}/n_{\rm vm}$ and $\delta$.  
} and
($iv$) the probability for an excited electron to undergo relaxation
on the contour, i.e., $\exp[-(D-2\delta)/l_0]$:
\begin{equation}
\mbox{HR} \sim 
h{\nu}{\cdot}2\pi\,(D-2\delta)\,
\sum\limits_{i=1}N^{(i)}_{\rm FWHM}\exp[-(D-2\delta)/l_0]\,.
\label{eq:HR_D}
\end{equation}
Here, the factor 2 accounts for two senses of propagation of vibration modes
along the closed contour, and the index $i$, without further elaborating
for the moment, identifies absorption / relaxation ``events'' likely to contribute
to the HR. This summation should, at least, take into account
different combinations ($m_{\rm el}$, $n_{\rm vm}$), within the global
limitation ${\leq}10$ imposed on these parameters, which retain
$m_{\rm el}/n_{\rm vm}$ and thus correspond to the same solution $D$
of Eq.~(\ref{eq:Dk_num}), but whose values $N_{\rm FWHM}$ are different.
In the next subsection, we'll adress an issue of multiple contours
which may participate in the absorption and the relaxation events,
in which relation the summation will be further explained.
For the time being, we retain the general structure of the expected expression
for the HR, and try to specify the relevant value of $l_0$.
The upper limit on it is the mean free path in bulk gold, ${\approx}\,35$~nm;
in nanoparticles, due to irregularities of internal structure,
$l_0$ ought to be much shorter. A priori, the smaller the size, the more likely
the crystal lattice is distorted, hence the smaller the mean free path. 
Estimates for some ``working'' value for small enough GNP sizes, which were covered
by the study by \cite{NanoRes2-400},
can be gained from their experimental data reproduced in Fig.~\ref{fig_01},
with the help of insight given by Eq.~(\ref{eq:HR_D}).

Fig.~\ref{fig_01} reveals a tendency towards the ``saturation'' of curves
with an increase of the gold volume fraction. This reflects the loss of efficiency
of heating the nanoparticles following their aggregation: the higher 
the GNP concentration, the higher the probability of their agglutination.
Therefore the ``net'' $l_0$ values are more safely to extract
from the slope of curves $d(\mbox{HR})/d(\mbox{volume fraction})$ near the origin.
Of interest for us is the value of $l_0$ for GNPs with $D_1=5$~nm
and $D_2=10$~nm; the corresponding slopes are almost identical.
Eq.~(\ref{eq:HR_D}) yields the following relation for the curves
corresponding to these values of $D$:
\begin{equation}
\frac{\mbox{HR}_2}{\mbox{HR}_1}\approx
\frac{D_2-2\delta}{D_1-2\delta}{\cdot}
\frac{\sum_i^{(D_2)} N^{(i)}_{\rm FWHM}}{\sum_j^{(D_1)} N^{(j)}_{\rm FWHM}}\,
\exp\left(\frac{D_1-D_2}{l_0}\right)\,.
\label{eq:Hr2Hr1}
\end{equation}
As mentioned above, each sum implies all relevant absorption / relaxation events 
within the particle of the corresponding size.
Practical calculations show that, in order to estimate
the $\mbox{HR}_2/\mbox{HR}_1$ ratio
from the data of Fig.~\ref{fig_01} with the accuracy of $\approx\,10$\%,
it suffices to retain in Eq.~(\ref{eq:Hr2Hr1}) the leading
term of each sum. Then, for the range of diameters $\simeq\,5-10$~nm,
$l_0$ can be expressed as follows:
\begin{equation}
l_0 = (D_1-D_2)\!\left[\ln\!\left(\frac{\mbox{HR}_2}{\mbox{HR}_1}{\cdot}
\frac{D_1-2\delta}{D_2-2\delta}{\cdot}
\frac{N^{(1)}_{\rm FWHM}}{N^{(2)}_{\rm FWHM}}
\right)\right]^{-1}\,.
\label{eq:l0_calc}
\end{equation}
The effective identity of the $d(\mbox{HR})/d(\mbox{volume fraction})$
slopes characterizing the GNPs of $D_1=5$~nm and $D_2=10$~nm implies
$\mbox{HR}_2/\mbox{HR}_1{\approx}1$ in Eq.~(\ref{eq:Hr2Hr1}). 
For the contour $\delta=0.7$~nm deep,
$(D_1-2\delta)/(D_2-2\delta)\,\approx\,0.42$. 
The relation $N^{(1)}_{\rm FWHM}/N^{(2)}_{\rm FWHM}$
follows from the straightforward counting of LAVMs that make a discrete spectrum
on a circuar contour of the $(D-2\delta)$ size: how many of modes will fall
within the FWHM of the longitudinal acoustic peak of gold.
In principle, $N_{\rm FWHM}$ steadily increases with size, 
but the subtlety is that not all steps in vibration energy
are compatible with the energy conservation criteria Eq.~(\ref{eq:mEl},
\ref{eq:m_el1}). 
Appendix \ref{sec:App2} explains this situation and argues that
$N^{(1)}_{\rm FWHM}=1$, $N^{(2)}_{\rm FWHM}\,\approx\,7.5$,
hence $N^{(1)}_{\rm FWHM}/N^{(2)}_{\rm FWHM}\,\approx\,0.13$, and 
from Eq.~(\ref{eq:l0_calc}) $l_0\,\approx\,-5\;\mbox{nm}/\ln(0.055) = 1.72$~nm.

Coming back to the discussion at this subsection's opening,
we can anticipate that the final HR, as function of GNP size,
will be the interplay (multiplication) of three tendencies.
The first one is the availablilty
of ``good'' $D$ values which can contribute at all; they group around $\approx$7~nm
and rapidly become scarse at larger sizes. The second tendency is the number
of LAVMs within the ``good'' energy interval (given by the elastic properties
of gold); this number essentially grows with $D$. The third effect is an
exponential cutting of the HR at the characteristic length  
much shorter than the electron mean free path in the bulk gold.
In total, the last tendency shifts the maximum of HR($D$) a bit more
to the left from the abovementioned ``primary'' $\approx\,7$~nm value 
than the second tendency shifts it to the right.

One can infer that, would the data for $D<5$~nm be available in Fig.~\ref{fig_01},
one could expect the initial slopes of the corresponding curves
(of HR vs gold volume fraction) to go steeper.  Correspondingly, the $l_0$ values
for such sizes would likely be less than 1.72~nm, and, on the scale of the GNP sizes,
the maximum of the HR would occur at slightly smaller values than
so far reported.

We'll see in the following how the final counting of HRs proceeds,
which also takes into account a somehow delicate issue of possible 
``diversification'' of the absorption / relaxation events.

\subsection{Case of multiple contours}

\begin{figure}[t]
\includegraphics[width=14.6cm]{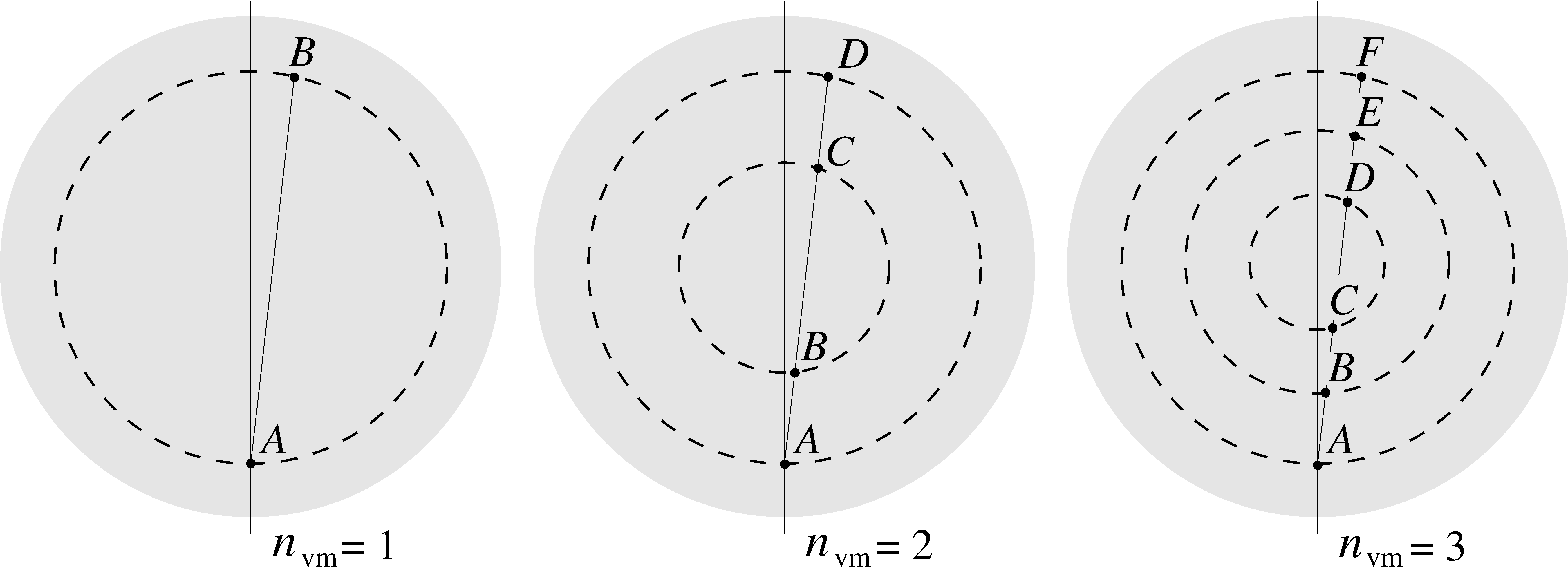}
\caption{Schematically drawn cyclic contours within nanoparticles
for three different values of the $n_{\rm vm}$ parameter.
``$A$'' marks the excitation site for a Fermi electron on the outer contour,
following the absorption of a RF photon and the LAVM. Other points indicate
possible sites of relaxation of the excited electron on traversing the particle.}
\label{fig_11}       
\end{figure}

At a risk of attributing too much precision to a simple enough model,
we would like to emphasize a possibility for the energy of excited electron
to be returned to a ``different'' phonon than that it was originally borrowed from.
Once $N_a$ and $m_{\rm el}$ are fixed for a GNP, 
the Eq.~(\ref{eq:m_el1}) can be, in principle, satisfied for various
values of $\delta'>\delta$, each $\delta'$ being a different ``depth'' of a 
circular contour, assuming for simplicity a symmetric placement of the latter
within the particle. The phonon energy at each contour is quantified, so that
\begin{equation}
(D-2\delta')/(D-2\delta)=n'_{\rm vm}/n_{\rm vm}\,,
\end{equation} 
for integer $n'_{\rm vm}$. Obviously, the case $n_{\rm vm}=1$ does not permit
any inner contour, whereas $n_{\rm vm}>1$ allows
$n'_{\rm vm}=1,\dots,n_{\rm vm}-1$. This is depicted in Fig.~\ref{fig_11}
for $n_{\rm vm}=1, 2, 3$,
and summarized in Table~\ref{tab:3}.
In this way, the discretized phonon momentum values (integer)${\times}h/L$
and the corresponding (momentum)${\times}v_{\rm L}$ energies must find
their ``resonance'' counterparts on relaxation. This is possible if
the ``secondary'' contour length is commensurate with the ``primary'' one.
The relation is not exact, since the energy absorbed by an electron is
that of phonon(s) \emph{plus} (much smaller) RF quantum, whereas the ``relaxation''
energy (released by the electron) is just the phonon(s)' one. However, it is helpful
for counting different contributions. 
An essential observation is that the density of ``resonance'' modes
decreases as the contour length shrinks. Still, the contributions from
``inner'' contours is not negligible; counting them, with different
$n_{\rm vm}/n'_{\rm vm}$ integer taken into account, would modify somehow
the contributions of different GNP sizes to the HR.

\begin{table}[b]
\caption{Characteristics of internal contours
and the electron path transversing them
for several small values of $n_{\rm vm}$.
See text for details and refer to Fig.~\ref{fig_11}.} 
\label{tab:3}
\smallskip
\begin{tabular}{c@{\hspace*{8mm}}c@{\hspace*{18mm}}c@{\hspace*{18mm}}c}
\hline\noalign{\smallskip}
$n_{\rm vm}$ & $n'_{\rm vm}$ & $(D-2\delta')/(D-2\delta)$ & $l/(D-2\delta)$ \\
\hline\noalign{\smallskip}
 2 & 1 & $1/2$ & $1/4^a$;~ $3/4^b$ \\ 
 3 & 1 & $1/3$ & $1/3^c$;~ $2/3^d$ \\ 
 3 & 2 & $2/3$ & $1/6^e$;~ $5/6^f$ \\
 4 & 1 & $1/4$ & $3/8$;~~   $5/8$ \\
 4 & 2 & $2/4$ & $2/8$;~~   $6/8$ \\
 4 & 3 & $3/4$ & $1/8$;~~   $7/8$ \\
\hline\noalign{\smallskip}
\end{tabular}\\
Fig.~\ref{fig_11}, middle panel: $^a\!AB$, $^b\!AC$. 
Fig.~\ref{fig_11}, right panel: $^c\!AC$, $^d\!AD$, $^e\!AB$, $^f\!AE$.
\end{table}

For each ``event'', the ``free path'' of an excited electron $l_i$
prior to its relaxation can be easily calculated in analogy to how it is shown,
for several selected ($n_{\rm vm}$, $n'_{\rm vm}$) combinations,
in the last column of Table~\ref{tab:3},
in terms of the contour size. The values are listed in pairs, indicating that, 
as the electron traverses the GNP approximately along the diameter
(by force of earlier envoked arguments), its ``exit point'' may occur
on a close or on a remote point on a given internal contour.

Alternatively, a situation is imaginable that the resonance criteria
of Eqs.~(\ref{eq:mEl}, \ref{eq:m_el1}) allow a modification of 
electron energy step on exchange with a phonon, on respecting the condition
\begin{equation}
\frac{\nu h}{m_{\rm el}{\cdot}\pi(D-2\delta)} =
\frac{\nu h}{m'_{\rm el}{\cdot}\pi(D-2\delta')}\,,\quad
\mbox{hence}\quad
\frac{D-2\delta'}{D-2\delta}=\frac{m_{\rm el}}{m'_{\rm el}}\,.
\end{equation}
Since the last relaton is $<1$ (the secondary contour at depth $\delta'$
is deeper than the primary one), $m'_{\rm el}>m_{\rm el}$; 
moreover, $m'_{\rm el}/m_{\rm el}$ must be an integer.

\begin{figure}[b]
\parbox[b]{8.8cm}{\includegraphics[width=6.6cm]{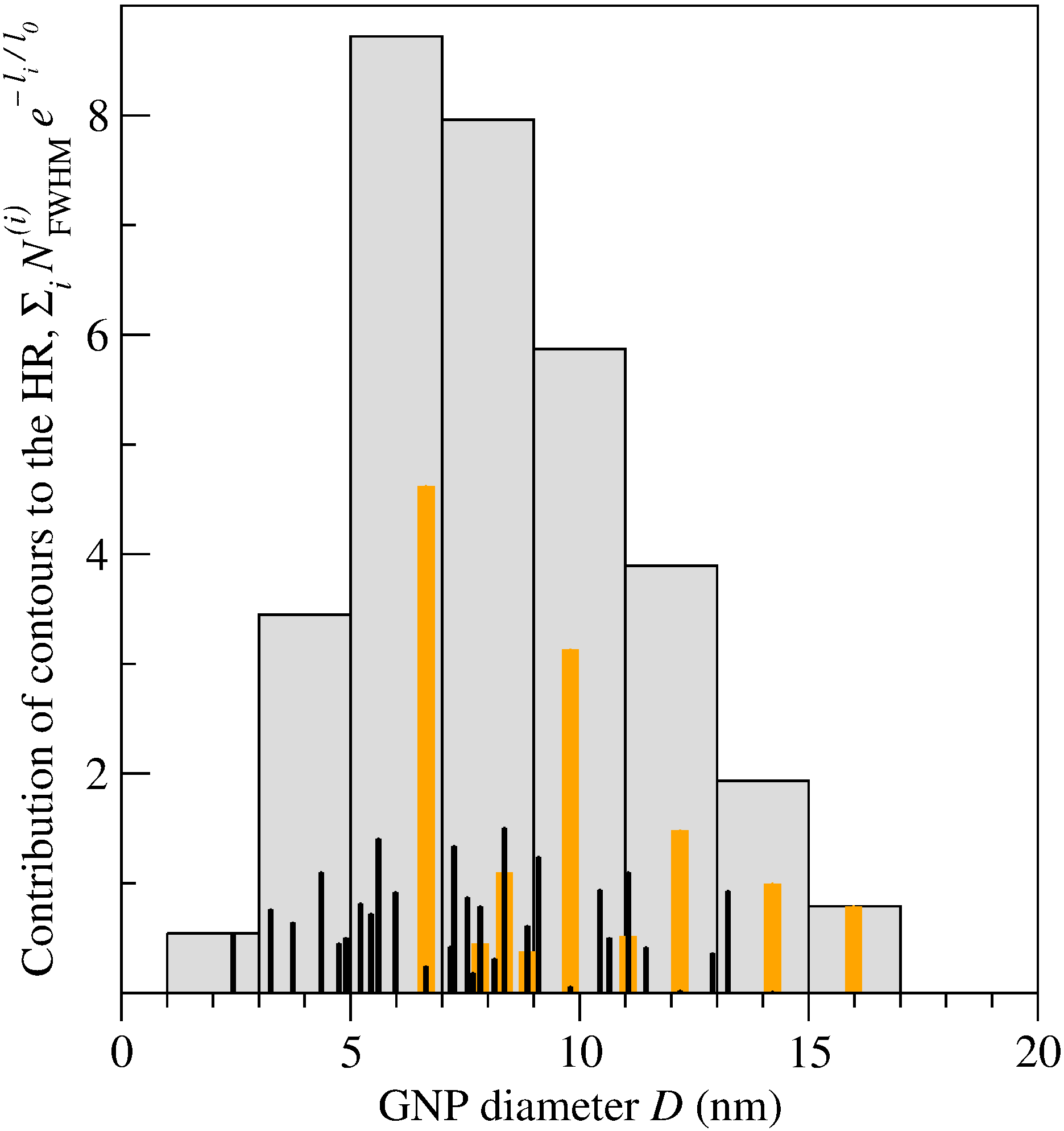}}
\parbox[b]{6.2cm}{
\caption{%
Contributions of possible contours in GNPs of different sizes
to their HR. Vertical lines mark the ``individual'' values of
$N^{(i)}_{\rm FWHM}\exp(-l_i/l_0)$ for each ``valid'' contour within
a ``resonant'' GNP size, whereby thin black lines stand for contributions of the
($m'_{\rm el}$ variable, $n_{\rm vm}=\mbox{const}$) type,
and thick orange lines -- for contributions of the 
($m_{\rm el}=\mbox{const}$, $n'_{\rm vm}$ variable) type.
Light gray bars make a histogram of the abovementioned contributions,
summed up within 2~nm steps of GNP diameter $D$.}
\label{fig_12}}     
\end{figure}

Possible contributions to the HR, in the spirit of Eq.~(\ref{eq:HR_D}),
are summarized in Table~\ref{tab:4} and depicted in Fig.~\ref{fig_12}.
The summation in the fourth column of Table~\ref{tab:4} selects the cases
($m_{\rm el}$ unchanged, $n'_{\rm vm}$ variable)
while that in the fifth column -- the cases
($m'_{\rm el}$ variable, $n_{\rm vm}$ unchanged).

\begin{table}[t]
\caption{%
Contributions to the HR from summations over primary and corresponding
secondary contours characterized by different ($m_{\rm el}$, $n_{\rm vm}$) values
in gold nanoparticles of diameters $D$. See text for details.}
\label{tab:4}
\smallskip
\begin{tabular}{%
r@{\hspace*{6mm}}c@{\hspace*{8mm}}r@{.}l@{\hspace*{10mm}}
l@{\hspace*{10mm}}l@{\hspace*{4mm}}}
\hline\noalign{\smallskip}
 & & \multicolumn{2}{c}{} & \multicolumn{2}{c}{\hspace*{0mm}\rule[-2mm]{0mm}{0mm}
 $\sum_i N^{(i)}_{\rm FWHM}\exp(-l_i/l_0)$} \\
 \cline{5-6}
$m_{\rm el}$ \hspace*{-2mm} & $n_{\rm vm}$ \hspace*{-2mm} &
\multicolumn{2}{c}{\hspace*{-8mm}$D$~(nm)} & 
$\sum_i{\equiv}\sum_{n'_{\rm vm}}, n'_{\rm vm}{\leq}10$ &
$\sum_i{\equiv}\sum_{m'_{\rm el}}, m_{\rm el}<m'_{\rm el}{\leq}10$
\rule[0mm]{0mm}{4mm} \\
\noalign{\smallskip}\hline\noalign{\smallskip}
  1 & 4 &  2&44 & $0.55$        & $0$ \\
  1 & 3 &  3&26 & $0.76$        & $0$ \\
  2 & 5 &  3&74 & $0.64$        & $0$ \\
  1 & 2 &  4&36 & $1.10$        & $0$ \\
  4 & 7 &  4&75 & $0.45$        & $0$ \\
  3 & 5 &  4&90 & $0.50$        & $0$ \\
  2 & 3 &  5&22 & $0.82$        & $0$ \\
  5 & 7 &  5&45 & $0.72$        & $0$ \\
  3 & 4 &  5&61 & $1.41$        & $0$ \\
  5 & 6 &  5&98 & $0.92$        & $0$ \\
  1 & 1 &  6&65 & $0.24$        & $4.62$ \\
  8 & 7 &  7&18 & $0.42$        & $0$ \\
  7 & 6 &  7&26 & $1.34$        & $0$ \\
  5 & 4 &  7&55 & $0.87$        & $0$ \\
  9 & 7 &  7&67 & $0.18$        & $0$ \\
  4 & 3 &  7&83 & $0.79$        & $0.45$ \\
 10 & 7 &  8&14 & $0.31$        & $0$ \\
  3 & 2 &  8&36 & $1.51$        & $1.10$ \\
  5 & 3 &  8&86 & $0.61$        & $0.38$ \\
  7 & 4 &  9&11 & $1.24$        & $0$ \\
  2 & 1 &  9&80 & $0.06$        & $3.13$ \\
  9 & 4 & 10&44 & $0.94$        & $0$ \\
  7 & 3 & 10&65 & $0.50$        & $0$ \\
  5 & 2 & 11&06 & $1.10$        & $0.52$ \\
  8 & 3 & 11&45 & $0.42$        & $0$ \\
  3 & 1 & 12&19 & $0.02$        & $1.48$ \\
 10 & 3 & 12&90 & $0.36$        & $0$ \\
  7 & 2 & 13&24 & $0.93$        & $0$ \\
  4 & 1 & 14&21 & $0.01$        & $1.00$ \\
  5 & 1 & 15&98 & $0$           & $0.79$ \\
  6 & 1 & 17&58 & $0$           & $0$ \\
  7 & 1 & 19&06 & $0$           & $0$ \\
  8 & 1 & 20&43 & $0$           & $0$ \\
  9 & 1 & 21&71 & $0$           & $0$ \\
 10 & 1 & 22&93 & $0$           & $0$ \\
\noalign{\smallskip}\hline\noalign{\smallskip}
\end{tabular}
\end{table}

The condition $m'_{\rm el}\,\leq\,10$ combined with that of 
$m'_{\rm el}/m_{\rm el}$ to be integer does considerably restrict the amount
of inner contours; hence so few non-zero entries in the fifth column
of Table~\ref{tab:4}. The largest contributions (for $D$=6.65 and 9.80~nm)
come out because, with their small initial $m_{\rm el}$ value,
the largest number of contours ($=4$) could be generated. 
The summation over possible contours and transitions yields a global scan
of the property $\sum_i N^{(i)}_{\rm FWHM}\exp(-l_i/l_0)$ which is expected
to be proportional to the HR, as function of the GNP diameter. The results
are shown in Fig.~\ref{fig_12} in two ways: as a contribution (weighted by 
$e^{-l_i/l_0}$) of each relaxation event, possible in a GNP of given size,
and as cumulate effect of multiple events, grouped in a histogram with a fixed
step in $D$. The first representation indicates that the contributions 
start from $D=2.44$~nm and drop down to zero beyond $D=15.98$~nm. The histogram
representation seems more helpful in inspecting the ``importance'' of different
GNP sizes for the HR. It seems that the HR efficiency peaks around 
$D\,\approx\,6$~nm and rapidly decreases for smaller and larger diameters.  

To conclude the discussion about multiple contours, we point out that
the excitation energy for an electron can also be borrowed from LAVM
propagating along one of internal contours. However, the probability
of such effects rapidly decreases with descending onto ever shorter contours
characterized by ever sparser distribution of quantized resonance modes.

\section{On the reduced heat production in aggregated GNPs}
\label{sec:4}
An aggregation of $n$ GNPs yields a larger particle, with the number of atoms
and electrons increased by the factor of $n$. However,
this won't be normally accompanied by a formation of joint
subsurface contour at the depth $\delta$, as the core regions
of different GNPs remain isolated from each other by their surface shells.
Therefore the ``optimal'' conditions of RF photon absorption, given by
Eqs. (\ref{eq:mEl}) and (\ref{eq:m_el1}), would be violated,
and the HR of aggregated GNPs reduced.

After Fig.~\ref{fig_01}, the HR eventually saturates, for all particle sizes,
as function of gold volume fraction. For small GNPs this saturation
occurs faster, for the apparent reason that, for a given gold volume fraction,
smaller particles mean their higher concentration, and hence higher
tendency for aggregation.

\section{Further suggestions for enhancing heating rates in GNPs}
\label{sec:5}
%
\begin{figure}[b]
\includegraphics[width=9.6cm]{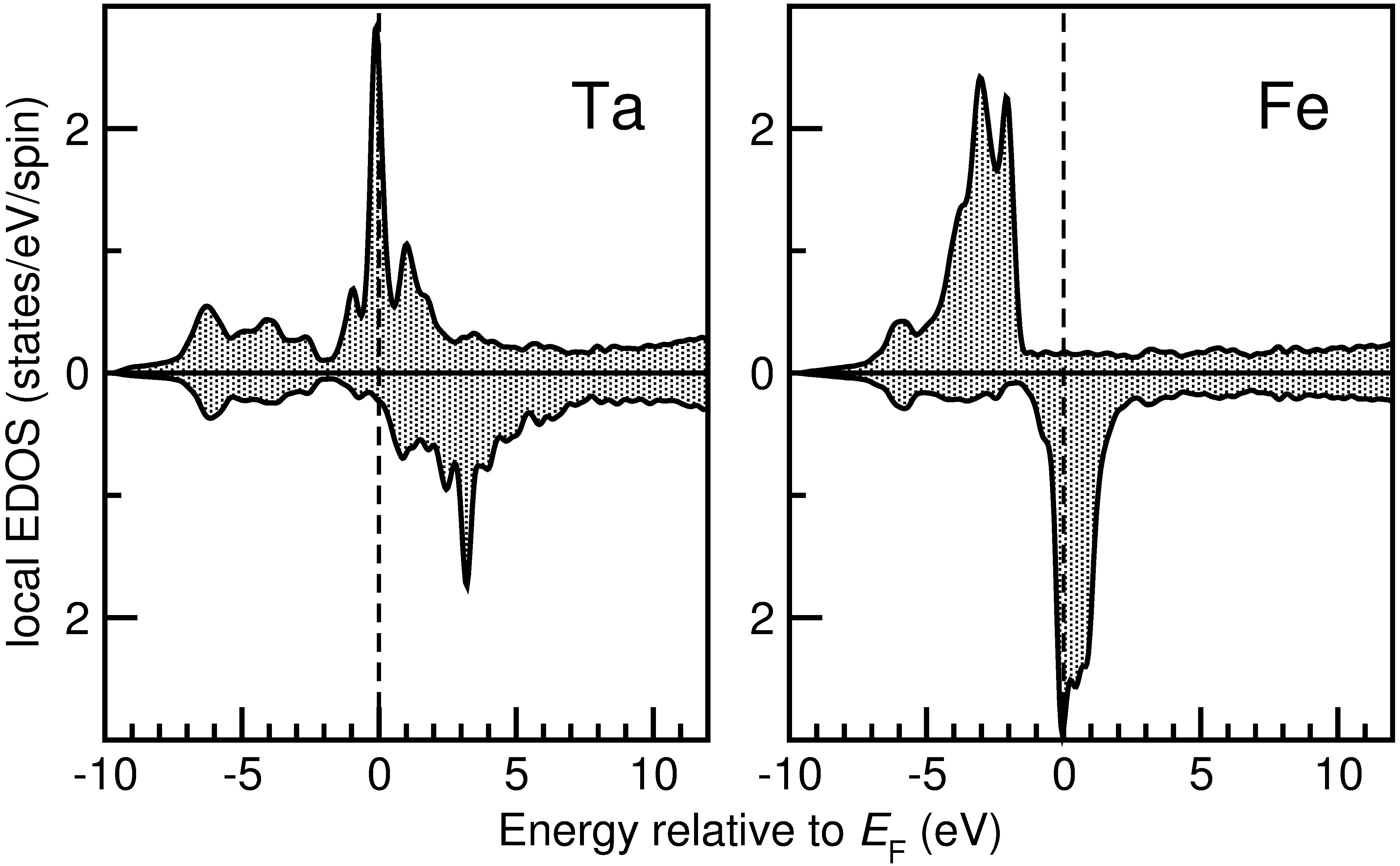}
\caption{Calculated spin-resolved local EDOSs against energy relative to $E_{\rm F}$
for Ta and Fe impurities in Au, reproduced from \cite{JNP6-061709}. Calculated
EDOS of \cite{PRB87-014301} for Fe $d$-electrons
in 32-atom quasirandom structure of Au$_{30}$Fe$_2$ supports our results
for Fe impurities in Au \citep{JNP6-061709}.}
\label{fig_13}       
\end{figure}

\noindent
In Sec.~\ref{sec:3}, it has been shown that in GNPs, the HR is enhanced
thanks to large number of participating LAVMs. As heating of GNPs involves
both the LAVMs and the Fermi electrons, one way of enhancing the HR ever further
would be to increase the number of electrons absorbing the RF photons.
This is possible through enhancing the local electronic DOS (EDOS)
at the Fermi energy of gold, doping the latter with transition metal impurities,
e.g., Ta or Fe -- see Fig.~\ref{fig_13}, and also \cite{JNP6-061709,patent_Mol+Pos}.
The main peaks of spin-split local EDOS
of these impurities, according to first-principles calculations,\footnote{%
Our calculations were done within the density functional theory,
using the generalised gradient approximation for the exchange-correlation,
by the {\sc Siesta} method
(see http://departments.icmab.es/leem/siesta/ ), 
allowing full atom relaxation for the 64-atom supercell 
(4$\times$4$\times$4-replicated fcc primitive cell with one atom
substituted by impurity).
}
are pinned at the Fermi level of gold, adding substantially to the bulk EDOS
of the latter.\footnote{%
The calculations done on 64-atom super cells (one impurity of Ta or Fe
per 63 Au atoms) indicate that the total EDOS at the Fermi level is increased
by ${\sim}50$\% compared to the pure Au.
}
In addition to just increasing the number of electrons absorbing the RF photons,
such doping would also enhance the scattering intensity of excited electrons.
Both of these tendencies are expected to result in enhanced HR.

Another imaginable way to bring about high EDOS would be due to $f$-electronic states
residing at the Fermi level, that is the case of heavy-fermion compounds -- 
see \cite{RMP56-755} for a review. Such compounds, typically possessing
a rare-earth element as their essential constituent, exhibit an anomalously high 
residual resistivity at low temperatures, traceable to high electron scattering 
intensity, and, in its turn, -- to a strong peak in the density of states of
$f$ electrons at the Fermi level. The presence of the compound
CeCu$_6$ among heavy-fermion systems
permits to presume that gold, like a homologue of copper, may host $f$-element
atoms so as to give rise to heavy-fermion behaviour. The nanoparticles
of such compounds might possess a yet elevated heating rate.    

\section{Discussion}
\label{sec:6}
The analysis of conservation conditions for energy
($m_{\rm el}{\cdot}{\Delta}E_{\rm el}\,\approx\, 
n_{\rm vm}{\cdot}{\Delta}E_{\rm vm}$) and momentum 
(i.e., for the movement direction of the excited electron
after absorption of the RF photon and the LAVM) lets estimate the size of GNPs
capable to produce high HR via a simultaneous involvement of several vibration modes
with energies matching $n_{\rm vm}{\cdot}{\Delta}E_{\rm vm}$, situated within
the LAVMs of gold. It follows from our estimates that the optimal size of GNPs
to use in RF hyperthermia is $5-7$~nm.
 
A number of known results seem to support our argumentation.
\cite{NanoRes2-400}, who explored heating of GNPs with sizes
from 5 to 250~nm in the electric field of 13.56~MHz frequency, observed that
the highest HR was inherent to GNPs of 5~nm size. 
\cite{IEEE_TransBiomedEng58-2002} heated the GNPs by the 13.56~MHz electric field
too, and, again, the smallest tested GNPs ($\approx\,5$~nm size) revealed
the maximum HR. 

Other researches testify that as the GNPs sizes deviate from $5-7$~nm,
the HRs change from bad to worse.
\cite{ADA535681} observed the heating of GNPs with the 4~nm
and 13~nm diameters and pointed out that their HRs were several times lower
than those for the 5~nm GNPs in the work by \cite{NanoRes2-400}.
\cite{JCollInterfSci358-47} exposed the GNPs with sizes of 20, 50
and 200~nm to the 13.56~MHz electric field and did not observe heating.
In view of our analysis, the GNPs of 20 or 50 nm fall short of sizes (shown
in Table~\ref{tab:1}) that would enable a ``direct'' absorption of a RF photon;
at the same time, these sizes are much larger than $5-7$~nm,
the ``optimum'' diameter for the phonon-assisted absorption.
As for the 200~nm size, that falls close to a number appearing in
Table~\ref{tab:1}, one can presume that the resonance 
for direct absorption is quite sharp and not satisfied
in the real GNPs studied.

\cite{Nanoscale4-3945} did not detect any significant heating
of GNPs with sizes of 15 to 30~nm (i.e. larger than the optimum size)
at the 13.56~MHz frequency. \cite{SovreTechnoMed4-30}
tried to treat tumours by the RF radiation of frequency 2.45~GHz using
the gold nanorods of 30~nm diameter and 60~nm length, whereby no anticancer
effect was identified. In our opinion, namely the large sizes of nanorods
were the problem. 

\cite{JPhysChemC116-24380} and \cite{Nanomed8-1096} pointed out
that heat generation was significantly reduced
when the GNPs were aggregated. We attribute it to the fact that, as the GNPs come
in contact, the common size of the formed aggregate is increased, and the condition
(\ref{eq:mEl}) is not anymore fulfilled, resulting in reduction of HRs of GNPs.

A valuable contribution to the analysis of experimental and theoretical studies
in heating kinetics of GNPs in the RF range was offered by a recent work
by \cite{Nanomed11-413}. The authors have noted
a spectacular failure of previous theoretical conclusions to account for
the experimental evidence that RF radiation is capable of heating the GNPs.
Our present work that emphasises the crucial role of LAVMs in the RF absorption
suggests a plausible explanation of the above problem.
 
\cite{ACSNano6-4483} who studied \emph{in vivo} the size
dependence (within the 2 to 15~nm range) of the GNPs' localization and penetration
in cancer cells, multicellular spheroids and tumours, concluded that the 2 and 6~nm
GNPs demonstrated advantages over larger nanoparticles in terms of tumour uptake
and permeability. Such GNPs were able to penetrate deeply into tumour tissue and
provide high levels of accumulation in it. As an application of the smallest 
(${\sim}2$~nm size) GNPs seems questionable due to loosening their biocompatibility
\citep{Small3-1941,MethEnzym509-225,Nanoscale5-6224},
the $5-7$~nm GNPs, possessing also the highest HRs, become very valuable
instruments in the RF hyperthermia.

Throughout the whole range of frequencies $\nu$ used in medical or biological
studies (${\nu}{\sim}10$~MHz -- 3~GHz), the RF photon energies $h\nu$ are small
compared to $n_{\rm vm}{\cdot}{\Delta}E_{\rm vm}$ of Eq.~(\ref{eq:mEl}),
for the LAVMs energy range
of ${\approx}3$ -- 21~ meV. Consequently, the condition (\ref{eq:mEl}) is
generally satisfied, and the GNPs of about $5-7$~nm size would have
elevated HRs
not only at 13.56~MHz, but everywhere within the mentioned interval of frequencies;
in particular, at 2.45~GHz, the operation frequency of a domestic microwave oven
with fine, inverter-regulated power control. The use of the latter for laboratory
investigations would permit performing genetic and biological experiments 
\emph{in vivo} -- e.g., with GNPs immersed in the tissues of larvae of such
classical research target as \emph{Drosophila melanogaster}.

Recently, \cite{IntJHypertherm29-99} reported that
at the frequency of 0.35~MHz the Pt nanoparticles possessed HR twice higher
than that of the GNPs and proposed to use them in RF hyperthermia. We attribute
high HR of Pt nanoparticles to enhanced EDOS at the Fermi level of Pt,
as compared to Au. For this reason, in Pt, a number of electrons able to absorb
the RF photons is enhanced. 

One should expect that in conditions of the experimental capacitive electric transfer 
system used by \cite{IntJHypertherm29-99} in nanoparticles
of AuPd and AuPt alloys
(Pd and Pt content within 40 -- 70 mass \%),
the HRs would be even higher than those in Pt nanoparticles. This is because
in these alloys, judging by their enhanced EDOS at $E_{\rm F}$,
the intensity of scattering of Fermi electrons is higher than that
in pure Pt. However, an application of AuPd, AuPt, Pd and Pt nanoparticles in the RF
hyperthermia is questionable due to their enhanced catalytic activities.

Compared with these nanoparticles, the GNPs containing impurities of Ta and/or Fe,
and hence developing an enhanced EDOS at the Fermi level and, expectedly,
an enhanced intensity of electron scattering, seem to be much more promising
for the RF hyperthermia using the GNPs \citep{JNP6-061709,patent_Mol+Pos}.  

During last years, the use of GNPs as systems of drug and gene delivery
into cancer cells has expanded considerably. An opportunity to synergistically
combine these techniques with RF hyperthermia may open new terrains
in contemporary cancer theranostics. Recently  
\cite{PNAS112-E1278}, apparently driven by a need to sense and overcome
the cancer multidrug resistance, invented an implantable hydrogel
with embedded DNA-coated GNPs of ${\sim}10$ -- 17~nm diameters.
Our above discussion hints that an attempt to go towards slightly smaller
particles, of $5-7$~nm diameter, would ``activate'' the embedded GNPs
for the RF hyperthermia, adding such an option to the GNPs' other functions.  

Mirkin and co-workers 
\citep{JACS134-1376,SciTranslMed5-209ra152,AnChInt54-527,GenDev29-732,
PNAS112-3892,PNAS112-5573} 
developed and used spherical nucleic acid 
GNP conjugates (13~nm diameter gold cores functionalized with densely packed
and highly oriented nucleic acids). Hypothetically,
the $5-7$~nm diameter GNPs along with optional RF heating thereof
could be used in these cancer treating technologies, providing new research
opportunities through temperature control.

Another issue that enters the domain of feasible is the 
transfer of technologies developed for the plasmonic heating of GNPs
onto the RF range.
Zharov and co-workers \citep{NatNanotech4-855,JBiophot6-523},
aiming to prevent metastasis, used magnetic trapping of tumour cells circulating
in the bloodstream with their simultaneous photoacoustic and photothermal detection.
To this end, the gold-plated carbon nanotubes were employed. 
As it seems, the use of the $5-7$~nm
diameter GNPs instead of gold-plated carbon nanotubes may help
to extend this method over applications in the RF range as well.

A recent work by \cite{PNAS112-1959} describes 
a promising so-called ``quantum rattle'', that is,
a hollow spherical particle (${\sim}150$~nm of total diameter)
with mesoporous silica shell (${\sim}25$~nm thickness) hosting both gold
quantum dots (AuQDs) of $<2$~nm diameter and GNPs (average crystallite size 7.3~nm
in diameter). The quantum rattle is highly biocompatible and combines both
cancer imaging and tumor treatment capabilities (chemotherapy and
photothermal therapy in near infrared range). All these advantages in matching
and even outperforming the state-of-the-art nanotechnology-based medical agents
are achieved thanks to just AuQDs, whereas a contribution of GNPs is very small.
Specifically, the photothermal therapy using the quantum rattles is provided through
excitation of AuQDs by an infrared laser, then the AuQDs emit infrared fluorescence
and enough heat to kill cancerous cells. In these quantum rattles, the GNPs are
likely just processing waste. Meanwhile, as their sizes (7.3~nm)
happen to fall close to the ``favourable'' range of $5-7$~nm,
the presence of GNPs in quantum rattles under discussion could be used
for additional targeted heating of the latter by RF irradiation.

Another recent work by \cite{EPJQuantTech2-19} outlines the use of gold
nanorods (of 10 nm diameter / 41 nm length) connected to nanodiamonds,
whereby the gold part serves to hyperthermia, by near-infrared laser heating,
and the nitrogen-vacancy centers in diamond serve for temperature sensing.
Here again, the use of nanoparticles of 5 -- 7 nm size in place of nanorods
would hopefully help to extend this technique over RF-induced hyperthermia. 

Last but not least, the RF heating of GNPs may be of interest for developers
of therapeutic strategies targeting to inhibit amyloidogenic process
in the Alzheimer's disease. \cite{NanoResLett3-435} suggested
to use in this context the GNPs of $12.5{\pm}1.7$~nm size
heated by the microwave 14~GHz radiation. Chances are that using the GNPs
of ``optimal'' sizes would be useful for these tasks, too.

\section{Conclusion}
\label{sec:7}
We suggest a physical model of the size effect in heat generation in GNPs,
which also accounts for a reduction of heat generation as the GNPs get aggregated.
In this model, the LAVMs (dominating in the distribution of vibrational density
of states) play an important role -- an apparently novel element in the related theory
framework. According to our model, the heating of GNPs is thought to consist
of two consecutive processes: first, the Fermi electron absorbs simultaneously
the RF photon and the LAVM available in the GNP; hereafter the excited electron
is relaxed, exciting a LAVM with the energy higher than that of the previously
absorbed LAVM. The model predicts that the GNPs to be effectively heated should
possess diameters of $\sim\,5-7$~nm, i.e., very close to the experimentally
inspected ${\sim}5$~nm. The absorption band is expected to be very wide
(${\sim}10$~MHz -- 3~GHz). This allows the use of frequencies typical for
the ``conventional'' RF hyperthermia (without conducting nanoparticles). The energy
release in the GNPs can be optimized by tuning the RF frequency, searching
a compromise between the HR energy transfer efficiency and a penetration depth
of the RF radiation into the biological tissue. The GNPs containing Ta or Fe impurity
atoms are expected to be more effective heaters compared to nanoparticles
of pure gold, due to enhanced electron density of states at the Fermi level.
Gold nanoparticles with rare-earth impurity atoms are also brought into consideration
as promising for the RF hyperthermia with conducting nanoparticles. The significance
of the present study follows from the fact that the cancer specialists seeking
approval for human clinical trials on the basis of their experimental results
remain thus far in the dark in what regards the physical mechanism
behind the observed trends.

\begin{acknowledgements}
The authors sincerely thank Drs. Curley, Mu{\~n}oz and Gong
for their kind permissions to use the figures from their works. We also would like
to thank Drs. Mu{\~n}oz and Kresch for useful discussions
of their measurements of the phonon DOS in gold.
\end{acknowledgements}

\appendix
\section{Level spacing vs number of atoms in a GNP}
\label{sec:App1}

\cite{JPSJ17-975,JPhysColloq38-C2-69} adopted the one-electron approximation
to electrons in small metallic particles, whereupon D.Y.~\cite{PRB63-193412}
predicted and J.~\cite{NatMater13-1007}
experimentally demonstrated the existence of the bulklike core region in them.
This justifies considering the GNP as a crystalline one and permits to specify
a relationship between the level spacing of electrons and number of atoms in GNP,
using formulas for the free electron gas.
A GNP of cuboid shape and consisting of $N_a$ atoms contains
(one ``free'' electron per atom of gold) $N_a$ electrons
filling the energy levels up to the Fermi energy $E_{\rm F}$.
The free-electron EDOS is \citep[e.g.,][]{wert1970physics}:
\begin{equation}
\rho(E) = dN'\!/dE = 4\pi(2m)^{3/2}\,L^3\,E^{1/2}\,h^{-3}\,,
\label{eq:A1}
\end{equation}
where $dN'$ is a number of states within the energy interval $dE$, $L$ is
the cuboid's edge and $h$ the Planck constant. Then the total number of electrons
is
\begin{equation}
N_a=\int\limits_0^{E_{\rm F}}\!\rho(E)\,dE=\frac{8\pi}{3}(2m\,E_{\rm F})^{3/2}
\left(\!\frac{L}{h}\!\right)^{\!\!3}\,.
\label{eq:A2}
\end{equation}
In accordance with Eq.~(\ref{eq:A1}), the spacing ${\Delta}E_{\rm el}$
between the Fermi level and nearest energy level above it
equals ${\Delta}E_{\rm el}\,\approx\,{\Delta}N'\!/\rho(E_{\rm F})$,
where ${\Delta}N'=2$, hence
\begin{equation}
{\Delta}E_{\rm el}\approx\frac{2}{\rho(E_{\rm F})}=\frac{h^3}
{2\pi(2m)^{3/2}L^3E_{\rm F}^{1/2}}\,.
\label{eq:A3}
\end{equation}
From Eqs.~(\ref{eq:A2}) and (\ref{eq:A3}), the Kubo's formula is deduced:
\begin{equation}
{\Delta}E_{\rm el}\,\approx\,\frac{4}{3}\,\frac{E_{\rm F}}{N_a}\,.
\label{eq:A4}
\end{equation}

\section{Counting vibration modes in GNPs of given size,
within given energy range}
\label{sec:App2}

Assuming that the LAVMs propagate with (longitudinal) sound velocity $v_{\rm L}$
along the circular contour at depth $\delta$ under the surface
of spherical particle of diameter $D$, the phonon energy quantum is
$\Delta E_{\rm vm}=v_{\rm L}h/[\pi(D-2\delta)]$. However, insofar as phonons
are absorbed /created in the process of the electron excitations / relaxations,
the phonon energies can only change in blocks, commensurable with the electron
states quantization, i.e., respecting the energy conservation condition
of Eq.~(\ref{eq:mEl}), 
$m_{\rm el}{\Delta}E_{\rm el}\,\approx\,n_{\rm vm}{\Delta}E_{\rm vm}$. 
Only certain combinations ($m_{\rm el}, n_{\rm vm}$) are possible, 
that selects the ``resonant'' values of $D$, as given by Eq.~(\ref{eq:Dk_num}).
Table \ref{tab:4} lists the allowed groups 
($m_{\rm el}$, $n_{\rm vm}$, $D$) in the increasing order of $D$.

Searching in Table~\ref{tab:4} for the $D$ values closest to the ``reference'' ones
$D_1=5$~nm and $D_2=10$~nm, one finds $D=5.22$~nm that comes along with
$n_{\rm vm}=3$, and $D=9.80$~nm that comes along with $n_{\rm vm}=1$.
We note that the apparently ``competitive'' values of $D$ by their closeness
to the reference values, namely, $D=4.90$~nm and 10.44~nm, can only be selected
with much higher values of $n_{\rm vm}$ and, consequently, may only
intervene with much sparser distribution, and hence much smaller impact,
or their allowed vibration modes. For $D=5.22$~nm,
$n_{\rm vm}{\Delta}E_{\rm vm}=3.35$~meV, so there is no more than 
just one vibration mode, of the energy $5{\times}(n_{\rm vm}{\Delta}E_{\rm vm})$,
that falls within the FWHM range of the interest, 14.6 to 18.4~meV (see caption to
Fig.~\ref{fig_08} and the related text), hence $N^{(1)}_{\rm FWHM}=1$
in Eq.~(\ref{eq:l0_calc}). For $D=9.80$~nm, 
$(n_{\rm vm}=1){\cdot}{\Delta}E_{\rm vm}=0.51$~meV,
and the above cited FWHM range may hosts much more modes at multipliers
of this energy, namely, $N^{(2)}_{\rm FWHM}{\approx}7.5$ 
in Eq.~(\ref{eq:l0_calc}).

\end{document}